\documentclass[10pt, conference, letterpaper]{IEEEtran}
\IEEEoverridecommandlockouts
\usepackage{cite}
\usepackage{amsmath,amssymb,amsfonts}
\usepackage{algorithmic}
\usepackage{algorithm}
\usepackage{graphicx}
\usepackage{textcomp}
\usepackage{xcolor}
\usepackage{diagbox}
\usepackage{graphicx}
\usepackage{float} 
\usepackage{subfigure}
\usepackage{xcolor}

\definecolor{gcolor}{RGB}{0,0,255} 
\definecolor{bcolor}{RGB}{0,0,0} 
\definecolor{ecolor}{RGB}{255,0,0}

\def\BibTeX{{\rm B\kern-.05em{\sc i\kern-.025em b}\kern-.08em
    T\kern-.1667em\lower.7ex\hbox{E}\kern-.125emX}}
\begin{document}






\title{EdgeVision: Towards Collaborative Video Analytics on Distributed Edges for Performance Maximization}

\author{
Guanyu Gao, Yuqi Dong, Ran Wang and Xin Zhou

\thanks{
G.Y. Gao and Y.Q. Dong are with School of Computer Science and Engineering, Nanjing University of Science and Technology (NJUST), Nanjing 210094, China. Email: \{gygao, dongyuqi\}@njust.edu.cn.
R. Wang is with College of Computer Science and Technology, Nanjing University of Aeronautics and Astronautics (NUAA), Nanjing 211106, China. Email: wangran@nuaa.edu.cn.
X. Zhou is with School of Communications and Electronics, Jiangxi Science and Technology Normal University (STNU), Nanchang 330027, China. Email: zhouxin@jxstnu.edu.cn. 
Corresponding~author:~Ran~Wang, Email: wangran@nuaa.edu.cn.
}}

\maketitle

\begin{abstract}
Deep Neural Network (DNN)-based video analytics significantly improves recognition accuracy in computer vision applications. Deploying DNN models at edge nodes, closer to end users, reduces inference delay and minimizes bandwidth costs. However, these resource-constrained edge nodes may experience substantial delays under heavy workloads, leading to imbalanced workload distribution. While previous efforts focused on optimizing hierarchical device-edge-cloud architectures or centralized clusters for video analytics, we propose addressing these challenges through collaborative distributed and autonomous edge nodes. Despite the intricate control involved, we introduce EdgeVision, a Multiagent Reinforcement Learning (MARL)-based framework for collaborative video analytics on distributed edges. EdgeVision enables edge nodes to autonomously learn policies for video preprocessing, model selection, and request dispatching. Our approach utilizes an actor-critic-based MARL algorithm enhanced with an attention mechanism to learn optimal policies. To validate EdgeVision, we construct a multi-edge testbed and conduct experiments with real-world datasets. Results demonstrate a performance enhancement of 33.6\% to 86.4\% compared to baseline methods.

\end{abstract}


\section{Introduction}
Video analytics plays a crucial role in various computer vision applications, such as video surveillance, augmented reality, and autonomous driving \cite{hu2014picwords}. Deep Neural Networks (DNNs) currently dominate the implementation of state-of-the-art video analytics algorithms. Despite their impressive accuracy, deploying DNN models for real-world video analytics presents challenges \cite{xiao2021towards}. DNN-based video analytics models often possess significant complexity with hundreds of layers, introducing computational overhead and substantial inference delays \cite{jiang2021joint,hao2022cdfkd}. Moreover, transmitting extensive video content to the cloud for inference incurs significant bandwidth costs and may result in intolerable transmission delays \cite{hu2018speeding}.

To enhance efficiency in video analytics applications by mitigating bandwidth costs and reducing data transmission delays, a viable strategy involves deploying DNN models on edge nodes situated in close proximity to end-users \cite{yi2017lavea, wu2023lean}. While these edge nodes can swiftly receive video data with minimal delay \cite{guan2020prefcache, cui2021towards}, their processing capacity is inherently constrained. In situations where a substantial influx of inference requests occurs, the edge node may become overwhelmed, exceeding its processing capabilities and resulting in pronounced inference delay. Concurrently, there are notable imbalances in the workloads of different edge nodes \cite{ma2017understanding}, with some nodes experiencing overload while others remain under-utilized. Therefore, a meticulous design of the video analytics mechanism integrated with edge computing is imperative to ensure optimal performance.

To enhance video analytics performance within the device-edge-cloud architecture, several studies (\cite{qian2022osmoticgate, ran2018deepdecision, hung2018videoedge, zhang2023device, wu2021edge, hao2022multi}) investigated offloading strategies. These efforts primarily concentrated on adapting video analytics configurations for preprocessing and model selection to strike a balance between inference accuracy and latency. In addressing bandwidth consumption, other studies (\cite{li2020reducto, li2021appealnet, wang2019bridging}) employed techniques such as frame filtering, inference difficulty classification, and resolution downsizing to reduce communication costs during video transmission while preserving recognition accuracy. Additionally, some works (\cite{jiang2018mainstream, jiang2018chameleon}) delved into optimizing configurations for video analytics.
Several studies (e.g., \cite{zeng2020distream}, \cite{jiang2021joint}, \cite{jingzong2023cross}, \cite{xu2023devit}) explored video analytics in scenarios involving multiple edge nodes or data centers. For instance, Distream \cite{zeng2020distream} focused on cross-camera workload balancing and partitioning between smart cameras and a centralized edge cluster. $A^2$ aimed at minimizing costs through dispatching inference requests among different data centers. CrossVision \cite{jingzong2023cross} and Polly \cite{jingzong2023cross} addressed the reduction of cross-camera content redundancy and workload balancing across edge cameras.

However, these efforts did not consider the joint control of multiple distributed edge nodes to maximize performance. Optimizing video analytics pipelines through edge collaboration presents significant challenges. Edge nodes must dynamically determine DNN model selection, video frame preprocessing configurations, and the optimal edge node for inference. These control parameters are interdependent, requiring a holistic approach considering the states and decision-making of all edge nodes to enhance overall performance. Moreover, in the scenarios where each edge node is autonomous, collaboration is essential while allowing each node to make independent decisions for received inference requests. Therefore, a distributed control mechanism is required to facilitate collaborative video analytics on the distributed edges.

We propose a MultiAgent Reinforcement Learning (MARL)-based approach for collaborative video analytics on distributed edges. Each edge node is treated as an autonomous agent, empowered to make distributed control decisions. Our algorithm employs a centralized training with a decentralized execution framework, enabling collaborative learning among agents to optimize overall performance. After training, an agent's control decision relies solely on its local state, mitigating communication costs. We integrate an attention mechanism to discern the importance of information collected from edge nodes, distilling valuable information for training. Performance evaluation involves a multi-edge video analytics testbed with extensive experiments using real-world datasets and settings. Experimental results demonstrate that EdgeVision significantly enhances performance by 33.6\%-86.4\% compared to baseline methods. The paper's main contributions are summarized as follows.

\begin{itemize}
    \item \emph{Design} a collaborative video analytics system facilitating edge nodes to collaborate in dynamic video preprocessing, model selection, and request dispatching to optimize system performance.

    \item \emph{Propose} an attention-based MARL approach for optimizing policy learning. Edge nodes collaborate to learn optimal policies, maximizing overall performance and executing decentralized control post-training.

    \item \emph{Evaluate} the effectiveness of our method using real-world datasets in diverse practical scenarios. Our approach showcases a significant improvement in overall performance, ranging from 33.6\% to 86.4\%, while concurrently achieving a remarkable 92.8\% reduction in video frame drop rates compared to baseline methods.
    
\end{itemize}

The paper is structured as follows. Section \ref{sec:related-work} provides an overview of related works, Section \ref{sec:system-design} details the system design and workflow, Section \ref{sec:problem-formulation} outlines the problem formulation, Section \ref{sec:algorithm} introduces the learning algorithm, and Section \ref{sec:experiment} demonstrates the experimental setup and results. The paper concludes in Section \ref{sec:conclusion}.

\section{Related Work} \label{sec:related-work}
This section reviews the existing works which focus on optimizing video analytics pipelines.

\textbf{Edge-to-cloud offloading.} Existing works explored video analytics performance optimization through edge-to-cloud offloading mechanisms. OsmoticGate \cite{qian2022osmoticgate} introduced a hierarchical queue-based offloading model to reduce real-time video analytics delay, considering resource and network capacity constraints. DeepDecision \cite{ran2018deepdecision} focused on client-side offloading, optimizing strategies by considering video compression, network conditions, and data usage. EdgeAdaptor \cite{zhao2022edgeadaptor} addressed accuracy, latency, and resource consumption trade-offs through model selection and application configuration. FastVA \cite{tan2021deep} studied computation offloading timing and Neural Processing Unit (NPU) usage on mobile devices to maximize accuracy while minimizing energy consumption. ModelIO \cite{wang2022dynamic} enhanced video analytics performance through dynamic model selection and resolution resizing. Zhang \textit{et al.} \cite{zhang2020decomposable} designed a Nash bargaining approach between edge and cloud for cooperative real-time video analytics computing. CEVAS \cite{zhang2021towards} adopted a collaborative edge-cloud framework dynamically partitioning video analytics pipelines. Rong \textit{et al.} \cite{rong2021scheduling} proposed an adaptive system for scheduling massive camera streams in an end-edge-cloud architecture. VideoEdge \cite{hung2018videoedge} designed a hierarchical edge-cloud video analytics framework considering multiple resources and accuracy trade-offs.

\textbf{Device-edge collaboration.}
Several studies explored the synergy between local devices and edge computing for video analytics. Long \textit{et al.} \cite{long2017edge} focused on efficiently partitioning video analytics tasks, distributing sub-tasks among edge nodes to boost performance. Lavea \cite{yi2017lavea} improved latency by optimizing task offloading and adjusting request priorities at the edges. Wang \textit{et al.} \cite{wang2020joint} proposed an online algorithm addressing configuration adaptation and bandwidth allocation for multiple video streams, optimizing the trade-off between accuracy and energy consumption in edge-based video analytics. FedVision \cite{deng2020fedvision} optimized end-to-end performance by provisioning video analytics workflows across devices and servers.

\textbf{Optimizing bandwidth consumption.}
Several studies focused on optimizing bandwidth cost for video analytics. DDS \cite{du2020server} employed a server-side DNN-driven approach maintaining high accuracy while minimizing bandwidth usage. CloudSeg \cite{wang2019bridging} reduced bandwidth cost by transmitting videos in lower resolutions and enhancing quality through super-resolution. SiEVE \cite{elgamal2020sieve} implemented semantic video coding to generate I-frames, avoiding decoding and analyzing the entire video. FilterForward \cite{canel2019scaling} and Reducto \cite{li2020reducto} targeted bandwidth reduction through edge-to-cloud and camera-based video frame filtering, respectively.

\begin{figure*}[h]
  \centering
  \includegraphics[width=0.8\linewidth]{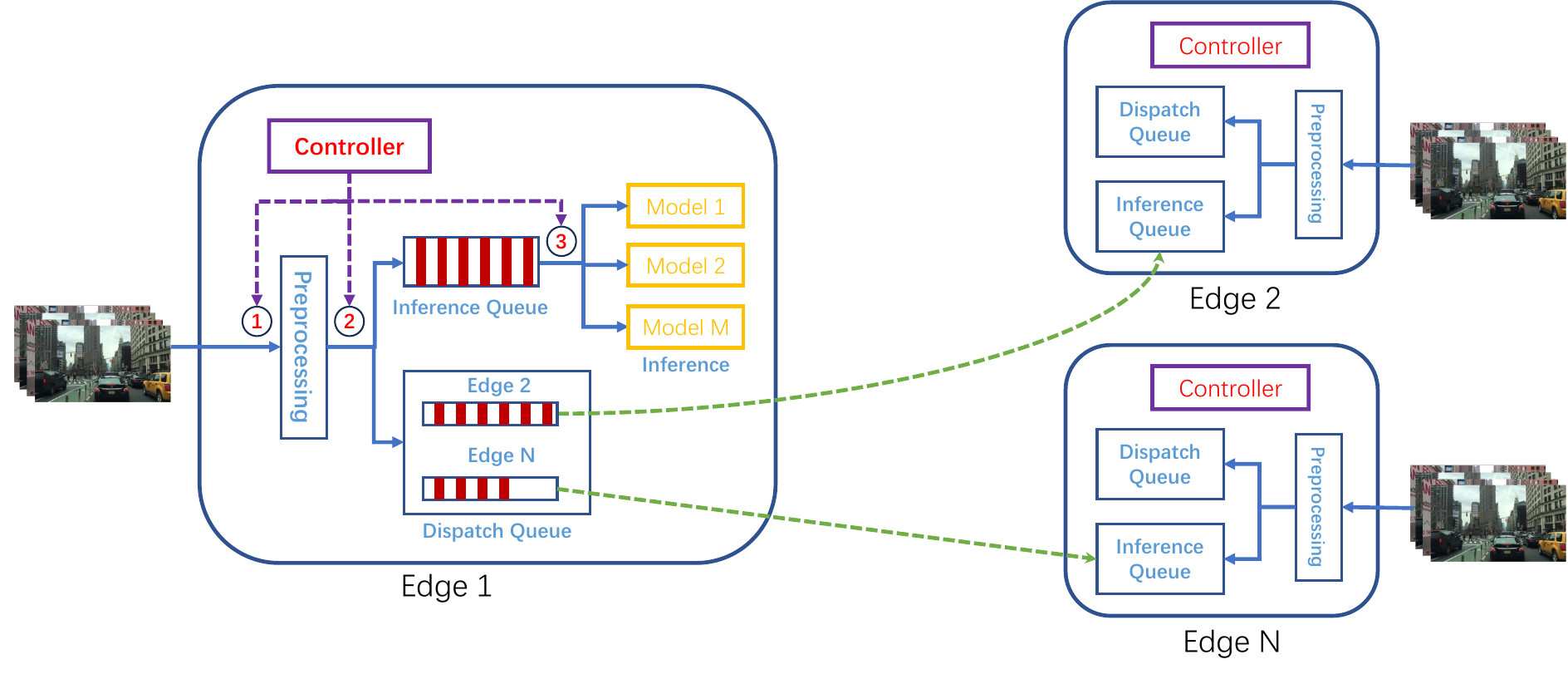}
  \caption{The system architecture of multi-edge collaborative video analytics. Multiple DNN models are deployed on each edge node, and the edge nodes jointly determine video preprocessing configurations, DNN models for inference, and inference locations to maximize the overall performances.}
  \label{fig:structure}
\end{figure*}

\textbf{Video analytics configuration.}
The system configurations for video analytics significantly impact resource consumption and accuracy. Mainstream \cite{jiang2018mainstream} dynamically adjusted resource sharing among applications for efficient resource utilization. Chameleon \cite{jiang2018chameleon} adapted configurations in real-time to optimize resource consumption and inference accuracy. Yoda \cite{xiao2021towards} developed performance clarity profiles for each video analytics pipeline, characterizing accuracy-to-cost tradeoffs. VideoStorm \cite{zhang2017live} optimized knob configurations for quality and delay in allocating resources for live video analytics. Zhao \emph{et. al} \cite{zhao2021reinforcement} proposed an RL-based method adjusting spatial and temporal settings for optimizing video analytics performance in terms of energy consumption and accuracy.

\textbf{Edge collaboration in video analytics.}
Several studies explored the collaboration of multiple edge nodes in video analytics. Distream \cite{zeng2020distream} introduced a distributed live video analytics system that dynamically balances and partitions camera cluster workloads, ensuring low latency, high throughput, and accuracy. However, it overlooked dynamic selection of DNN models, video preprocessing configurations, and time-varying edge node bandwidth. $A^2$ \cite{jiang2021joint} strategically placed compressed DNN models at various geo-locations, optimizing inference requests for different model versions. Nevertheless, it did not consider request dispatching among different data centers. Polly \cite{jingzong2023cross} facilitated co-located edge cameras with overlapping views to share inference results, reducing redundant object detection and enhancing resource efficiency. CrossVision \cite{zhang2023crossvision} designed a distributed framework for real-time video analytics, minimizing inference delay and maximizing accuracy by discovering video content correlation and balancing workloads among cameras. However, these works concentrated on mitigating redundant video content overlap across different cameras.

\section{System Overview} \label{sec:system-design}
This section presents the architecture and workflows for collaborative video analytics on distributed edges.

\subsection{System Architecture}
The system architecture for multi-edge collaborative video analytics is depicted in Fig. \ref{fig:structure}. Geographically distributed edge nodes collaborate to handle video analytics inference requests from specific regions. Each edge node is deployed with multiple DNN models for inference, each with distinct profiles in terms of accuracy and delay. Larger, more complex models offer higher accuracy but entail longer inference delays compared to smaller, simpler models.

Upon receiving an inference request, the edge node can $\textcircled{1}$ resize the video frame to a lower resolution to reduce transmission and inference delay. The edge node can then either $\textcircled{2}$ perform local inference or dispatch the request to another edge node. For local inference, $\textcircled{3}$ the preprocessed video frame is placed in the task queue and awaits inference using a chosen DNN model. If the local edge node is overloaded, the request can be dispatched to another edge node. The preprocessed video frame is placed in the corresponding dispatch queue and forwarded to a specified remote edge node for inference using a selected DNN model.

\subsection{Distributed Control with Edge Collaboration}
In multi-edge collaborative video analytics, control decisions at each edge node, such as DNN model selection, inference node selection, and video preprocessing, significantly impact the overall system performance. These decisions are influenced by factors like request rate, local pending tasks, and bandwidth between edge nodes. The control policies of the agents mutually influence each other, necessitating collaborative learning to maximize overall system performance.

Unlike a centralized controller, we adopt a distributed control scenario, where each edge node autonomously makes decisions to optimize overall performance. A centralized controller relies on information from each edge node, which may not always be timely available due to privacy concerns or communication delays. Additionally, it introduces increased communication burdens for decision-making based on the information of all edge nodes.

\begin{table}[h]
\caption{Key Notations}
\centering
\begin{tabular}{p{0.7cm} p{7cm}}
\hline
\hline
$t$ & Time slot index, $t=0,1,2,...$\\
$N$ & Total number of edge nodes/agents\\
$i$ & Edge node/agent index \\
$\lambda_{i}(t)$ & Request arrival rate on edge node $i$ at time slot $t$ \\
$\Upsilon^i_t$ & Request arriving on edge node $i$ at time slot $t$ \\
$l_{i}(t)$ & Inference task queue length at edge node $i$ during time slot $t$ \\
$q_{ij}(t)$ & Dispatch queue length from edge node $i$ to $j$ at time $t$ \\
$b_{ij}(t)$ & Bandwidth between edge nodes $i$ and $j$ during time slot $t$\\
\hline
$e$ & Selected edge node for inference\\
$m$ & Selected DNN model for inference\\
$v$ & Selected resolution for video preprocessing\\
$\mathcal{E}$ & Set of candidate edge nodes\\
$\mathcal{M}$ & Set of candidate DNN models\\
$\mathcal{V}$ & Set of candidate video resolutions\\
$D_v$ & Average delay for downsizing a video frame to resolution $v$ \\
$B_{v}$ & Data size of a video frame at resolution $v$ \\
$I_{m,v}$ & Inference time with selected model $m$ and resolution $v$\\
$P_{m,v}$ & Accuracy with selected model $m$ and resolution $v$\\
\hline
$q_t^i$ & Local queuing delay for the request on edge node $i$ at $t$\\
$d_t^i$ & Overall delay for the request arriving on edge node $i$ at $t$\\
$P_{i}(t)$ & Set of requests completed on edge node $i$ during time slot $t$\\
$\omega$ & Penalty weight for the overall delay\\
$T$ & Frame drop time threshold\\
$F$ & Penalty for frame dropping\\
$\chi_t^i$ & Cost for the request arriving on edge node $i$ at time slot $t$\\
\hline
$s(t)$ & Global state of the environment at time slot $t$ \\
$o_{i}(t)$ & Local state of edge node $i$ at time slot $t$\\
$a_{i}(t)$ & Control action for edge node $i$ at time slot $t$\\
$r_i(t)$ & Reward for edge node $i$ at time slot $t$\\
$r(t)$ & Shared reward at time slot $t$\\
$\pi_{i}^{*}$ & Optimal control policy for edge node $i$\\
$\gamma$ & Discount factor for the shared reward\\
\hline
$\mu_{i,\theta}$ & Actor network of agent $i$ under the parameters $\theta$\\
$V_{i,\phi}$ & Critic network of agent $i$ under the parameters $\phi$\\
\hline
\hline
\end{tabular} \label{tab:notations}
\end{table}

\section{System Model and Problem Formulation} \label{sec:problem-formulation}
This section introduces system models for collaborative video analytics across multiple edges and formulates the performance maximization problem using multiagent reinforcement learning. The key notations are summarized in Table \ref{tab:notations}.

\subsection{Inference Request Arrival}
We model the discrete time system as $t=0,1,2,...$ with each time slot lasting a short duration (e.g., 100ms), typically accommodating only one or zero request.
These requests encompass images from users, video frames from cameras, or data from other devices seeking recognition, all potentially reaching an edge node.
Assuming a system of $N$ homogeneous edge nodes, the arrival rate of inference requests on each edge node is non-stationary and follows an arbitrary distribution.
We represent the arrival rate of inference requests on edge node $i$ at time slot $t$ as $\lambda_{i}(t)$, and denote the inference request arriving at time slot $t$ on edge node $i$ as $\Upsilon^i_t$.

\subsection{Video Preprocessing}
Prior to inference or transmission to other edge nodes, video frames undergo preprocessing, including operations such as resizing, cropping, and quality downsizing. This work specifically investigates resolution resizing as a case study.
For an inference request, the video resolution can be reduced to a lower resolution $v$ from the set $\mathcal{V}$ \cite{zhang2020deepqoe}, encompassing available resolutions like 360P, 480P, 720P, etc. We assume an average delay of $D_v$ for downsizing a video frame to resolution $v$, with the corresponding data size denoted as $B_{v}$.

\subsection{Model Selection}
The set of deployed DNN models on individual edge nodes is denoted as $\mathcal{M}$, assuming uniform computing capacities across edge nodes. The inference time for a video frame is influenced by both the chosen DNN model's complexity (represented by $m$) and the video frame resolution (denoted as $v$). We define the inference time for a video frame as $I_{m,v}$ and the corresponding accuracy as $P_{m,v}$.

\subsection{Local Inference}
The inference task queue length of edge node $i$ at time slot $t$ is denoted as $l_{i}(t)$. The queuing delay for local inference is determined by the pending requests in the local edge node's task queue and the inference time of each request.

For a locally processed request $\Upsilon^i_t$, the queuing delay is calculated as follows:
\begin{equation}
q^i_t = \sum^{l_{i}(t)}_{\tau=0} I_{m_\tau^i,v_\tau^i},
\end{equation}
where $I_{m_\tau^i,v_\tau^i}$ is the inference time of the $\tau$-th task in the task queue of edge node $i$ with model $m_\tau^i$ and resolution $v_\tau^i$. To prevent system overload and ensure real-time inference for video frames, requests exceeding a queuing time threshold are dropped from the queue.

If request $\Upsilon^i_t$ is successfully completed, the overall delay for local inference is given by:
\begin{equation}
d_t^i = D_{v_t^i} + q_t^i + I_{m_t^i,v_t^i},
\end{equation}
where $D_{v_t^i}$ and $I_{m_t^i,v_t^i}$ represent the pre-processing time and inference time for request $\Upsilon^i_t$, respectively.

\subsection{Remote Inference}
Let $q_{ij}(t)$ denote the dispatch queue length from edge node $i$ to edge node $j$ at time slot $t$. The queuing delay in the dispatch queue is influenced by the transmission time of tasks from edge node $i$ to edge node $j$. If request $\Upsilon^i_t$ is dispatched for inference to edge node $j$, the queuing delay in the dispatch queue is given by:
\begin{equation}
f_t^i = \sum^{q_{ij}(t)}_{\tau=0} \frac{B_{v^{ij}_\tau}}{b_{ij}(\tau)},
\end{equation}
Here, $B_{v^{ij}_\tau}$ represents the data size of the $\tau$-th task in the dispatch queue of edge node $i$ to edge node $j$, and $b_{ij}(\tau)$ is the bandwidth for transmitting the $\tau$-th task.

The overall delay for the request is calculated as:
\begin{equation}
d_t^i = D_{v_t^i} + f_t^i + \frac{B_{v^i_t}}{b'_{ij}} + \sum^{l_j(t')}_{\tau'=0} I_{m_{\tau'}^j,v_{\tau'}^j} + I_{m_t^i,v_t^i},
\end{equation}
where $B_{v^i_t}$ is the data size of request $\Upsilon^i_t$ after preprocessing, $b'_{ij}$ is the bandwidth for transmitting the video frame, and $l_j(t')$ is the length of the task queue on edge node $j$ when it receives the request at time slot $t'$, where $t'=t+f_t^i+\frac{B_{v^i_t}}{b'_{ij}}$.

\subsection{System Performance}
The system's performance is evaluated using two metrics: recognition accuracy and overall delay. To account for varying priorities, we define system performance as a weighted sum of these metrics. If a request is completed successfully, the performance is a linear combination of accuracy and delay. In situations of system overload, causing intolerable delays, a request will be dropped if the queuing delay exceeds a threshold. A dropped request incurs a fixed penalty. The performance for request $\Upsilon^i_t$ is calculated as:
\begin{equation}
\chi^i_t = \left\{
\begin{array}{cl}
P_{m_t^i,v_t^i} - \omega*d^i_t, & d^i_t \le T, \\
-\omega*F,  & d^i_t > T,\\
\end{array} 
\right.
\end{equation}
Here, $P_{m_t^i,v_t^i}$ is the recognition accuracy of the request, $F$ is a constant, and $\omega$ is the penalty weight for the delay. A higher penalty weight indicates increased significance assigned to the overall delay in the inference of a video frame.

\subsection{Problem Formulation}
We cast the problem of maximizing performance in multi-edge collaborative video analytics as a multiagent reinforcement learning problem. Each edge node is treated as an agent collaborating to learn the optimal policy for maximizing overall system performance.

\textbf{State:}
Each edge node can observe its local state, including the history inference request rate ($\lambda_{i}(t)$), the length of the local task queue ($l_{i}(t)$), the lengths of the dispatch queues to the other edge nodes ($q_{ij}(t)$), the bandwidths between the local edge node and the other edge nodes ($b_{ij}(t)$).
We denote the local state of edge node $i$ at time slot $t$ as: 
\begin{equation} \label{eq:local}
o_{i}(t) = (\lambda_{i}(t), l_{i}(t), q_{ij}(t), b_{ij}(t)).
\end{equation}

The multi-edge environment's global state consists of the local observations of each agent. The global state at time slot $t$ is denoted as:
\begin{equation} \label{eq:state}
s(t) = (o_1(t), o_2(t), ..., o_N(t)).
\end{equation}

\textbf{Action:} 
Each edge node performs control actions for each video frame to conduct video analytics. These actions are composed of the parameters for the inference request, determining the chosen edge node for inference (either local or dispatched to another edge node), the specified DNN model, and the desired resolution for video preprocessing.
We denote the action for edge node $i$ at time slot $t$ as: 
\begin{equation}
a_{i}(t) = (e, m, v), e \in \mathcal{E}, m \in \mathcal{M}, v \in \mathcal{V},
\label{eq:action}
\end{equation}
where $e$ is the chosen edge node for inference and $\mathcal{E}$ is the set of edge nodes.
If the selected edge node for inference matches the edge node that receives the inference request, the inference will be conducted locally. However, if the selected edge node differs from the receiving edge node, the request will be dispatched to the corresponding edge node for inference.

\textbf{Reward:} 
We define the reward for edge node $i$ at time slot $t$, denoted as $r_i(t)$, as the sum of the system performance of the inference requests completed on edge node $i$ during time slot $t$.
We denote the set of inference requests completed on edge node $i$ at time slot $t$ as $P_{i}(t)$, and the reward $r_i(t)$ can be calculated as:
\begin{equation}
   r_i(t) = \sum_{p \in P_i(t)} \chi_p,
\end{equation}
where $\chi_p$ is the quantitative performance of the $p$-th request in $P_i(t)$.  
Because the edge nodes work cooperatively to optimize the overall performance, we design the reward function as a shared reward, which is the sum of the reward of all edge nodes. The shared reward is calculated as: 
\begin{equation}
\label{eq: share reward}
   r(t) = \sum_{i=1}^{N} r_i(t).
\end{equation}

\textbf{Optimization objective:}
At time slot $t$, agent $i$ observes local states $o_{i}(t)$ and executes control action $a_{i}(t)$ following control policy $\pi_{i}$.
The system state transitions from $s(t)$ to $s(t+1)$ through joint control actions of agents and time-varying workloads. 
At the time slot's end, agent $i$ observes its reward $r_i(t)$ and the shared reward of all agents, $r(t)$. 
Given diverse workload distributions across edges, the objective is to learn each agent's optimal policy $\pi_i^*$ to maximize discounted overall shared reward, which can be presented as:
\begin{equation}
\pi^* = \arg\max \mathbb{E}_{s(t),a(t)}[\sum_{k=0}^{\infty}\gamma^{k}r(t+k)].
\end{equation}
Here, $\pi^*$ denotes the set of optimal policies for the agents, and $r(t + k)$ represents the shared reward at time slot $t + k$.

\section{Algorithm Design} \label{sec:algorithm}
This section outlines the design rationales behind our algorithm and introduces the network architecture, along with the multiagent policy updating mechanism.

\subsection{Design Rationale}
We design our learning algorithm with the following principles for multi-edge collaborative video analytics.

\textbf{1) Collaborative learning:} The control policies of the agents influence each other. This necessitates a fully cooperative approach among the edge nodes to jointly learn their optimal policies and maximize the collective reward.

\textbf{2) Distributed control:} During training, agents exchange information about other agents' states, which are not locally observable, to facilitate collaborative learning. After training, the agents can make decisions based solely on their own local states. This enables distributed control of the edge nodes and significantly reduces communication overhead, and it is particularly effective in scenarios where each edge node functions as an autonomous entity.

\textbf{3) Knowledge distillation:} As the number of edge nodes increases, the state space also grows, posing challenges for individual agents to learn optimal policies effectively. The increasing irrelevant information from other agents can hinder an agent's learning performance. To address this, we leverage the attention mechanism to distill valuable information and enhance the performance of collaborative learning.

\begin{figure*}[h]
  \centering
  \includegraphics[width=\linewidth]{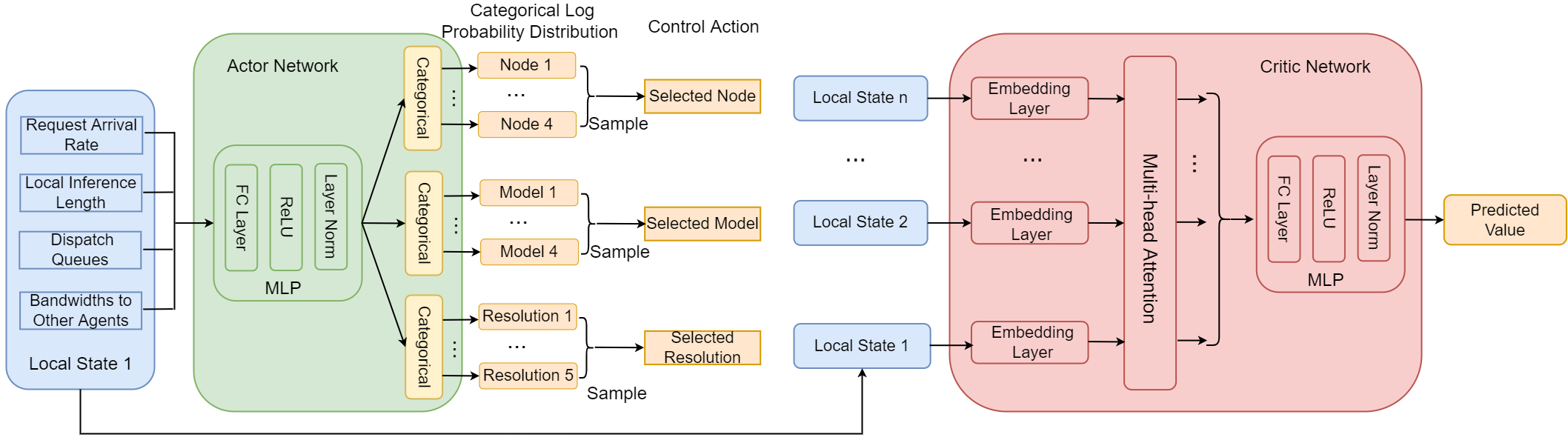}
  \caption{The network structure of the actor and critic networks of an agent. The actor network makes control decisions solely based on the local state, while the critic network makes predictions based on global states.}
  \label{fig:network_attention}
\end{figure*}

\subsection{Network Architecture}
\textbf{Actor-critic framework:}
In our approach, each edge node acts as an independent agent, equipped with a dedicated actor network and a critic network.
The critic network is enabled during the training phase, but once the training is completed, each agent only relies on its actor network to make control actions.
After training, an agent's actor network operates independently and does not depend on the state information of other agents.
In Fig. \ref{fig:network_attention}, we present the network structures of both the actor and critic networks. 
The actor network for agent $i$ with parameters $\theta$ is represented by $\mu_{i,\theta}(o_{i}(t))$, whereas the critic network with parameters $\phi$ is denoted as $V_{i,\phi}(s(t))$. Here, $o_{i}(t)$ represents the local state of agent $i$ at time slot $t$, and $s(t)$ denotes the global state at that time.
The actor network generates actions by mapping the local state defined in Eq. \eqref{eq:local} to multiple discrete action probability distributions, as defined in Eq. \eqref{eq:action}. The agent can then select discrete control actions by sampling from these probability distributions.
On the other hand, the critic network serves as a value function. It takes as input the local states of all agents (i.e., the global state) and produces predicted values.

\textbf{Attentive critic network:}
As the number of edge nodes increases, the input dimension of the critic network grows.
However, not all edge nodes affect each other, and some nodes far apart may have few connections.
On the other hand, the larger dimensions of the state spaces may potentially lead to larger prediction errors due to irrelevant information. Consequently, this can impact the actor network's ability to learn the optimal policy. To avoid indiscriminate attention to all agents' information, we incorporate the multi-head attention mechanism to distill valuable information.

1) Local state embedding: The embedding layer is comprised of a single Multi-Layer Perceptron (MLP) network. Each agent is equipped with its own dedicated embedding network for a critic network, denoted as $\Theta_{i}$. The embedding process for agent $i$ is expressed as
\begin{equation}
e_{i} = \Theta_{i}(o_{i}),
\end{equation}
where $o_{i}$ represents the local state of agent $i$, and $e_{i}$ is the output of the corresponding embedding network.

2) Attention Network:
The embedding networks generate multiple vectors, subsequently input into the multi-head attention network. This network efficiently captures crucial information during training, bolstering the critic network's precision in value predictions and enhancing overall system performance. We denote the multi-head attention network for an agent as $\Psi$, expressed by the equation:
\begin{equation}
(\psi_{1}, \psi_{2}, ..., \psi_{n}) = \Psi(e_{1},e_{2}, ...,e_{n}),
\end{equation}
where $e_{1}, e_{2}, ..., e_{n}$ represent the outputs of the embedding layers. The resulting $\psi_{1}, \psi_{2}, ..., \psi_{n}$ signify the multi-head attention network outputs. This attention network functions by mapping from a query and a set of key-value pairs to a weighted sum of values.

3) Attentive critics:
Each agent is equipped with an individual critic network, denoted as $f$, which comprises a final two-layer MLP network. The multi-head attention network generates outputs that are concatenated and subsequently fed into the final two-layer MLP network. This process yields the predicted value, represented as
\begin{equation}
\label{eq: predicted value}
    v = f(\psi_{1}, \psi_{2}, ..., \psi_{n}),
\end{equation}
where $v$ signifies the predicted value for an agent, serving as the output of the overall critic network.

\subsection{Training Methodology}
We now demonstrate the training process for the actor and critic networks of each agent using the proposed algorithm. The specific training procedures are outlined in Algorithm \ref{alg:algo}.

At each time slot $t$, agent $i$ observes local working state $o_{i}(t)$ and communicates to acquire global state $s(t)$. The local state $o_{i}(t)$ is input to the actor network $\mu_{i,\theta}$ to generate a categorical log probability distribution for actions. Subsequently, actions $a_{i}(t)$ are sampled from these distributions, 
\begin{equation}
\label{eq: sample action}
p_{i}(t) = \mu_{i,\theta}(o_{i}(t)),
a_{i}(t) \sim p_{i}(t),
\end{equation}
where $p_{i}(t)$ represents the categorical log probability distributions for agent $i$ at time slot $t$. Following this, each agent conducts video analytics for the inference requests in accordance with the control actions.
At the end of time slot $t$, each agent computes its reward $r_i(t)$ for the given time slot. The shared reward $r(t)$ for time slot $t$ is then calculated according to Eq. \eqref{eq: share reward}. Agent $i$ obtains its new local state $o_{i}(t+1)$ and the global state $s(t+1)$ from the video analytics system at the onset of time slot $t+1$.
The local state, global state, action, shared reward, new local state, and new global state are stored as a transition in the experience buffer, represented as $(o_{i}(t), s(t), a_{i}(t), r(t), o_{i}(t+1), s(t+1))$.

After an episode, the estimated advantage is calculated with Generalized Advantage Estimation (GAE) \cite{schulman2015high} in trajectory $\tau$. 
The output of the critic network is the predicted value $V_{\phi}(s(t))$ and the joint advantage function $ \hat{A}(s_t, a_t) $ is estimated through a truncated version of GAE. 
The estimation of the advantage value $\hat{A}_t$ at time slot $t$ is as follow,
\begin{equation}
\label{eq: GAE}
    \hat{A}_t = \hat{A}(s_t, a_t) = \delta_t + (\gamma \lambda) \delta_{t+1} + \ldots + (\gamma \lambda)^{T-t-1} \delta_{T-1},
\end{equation}
where $\delta_t = r(t) + \gamma V_{\phi}(s(t+1)) - V_{\phi}(s(t))$. 
By adjusting the hyperparameter $\lambda$, we can reduce variance while maintaining an acceptable level of bias.
The reward-to-go $\hat{R}_t$ at time slot $t$ in the trajectory is calculated as follow,
\begin{equation}
\label{eq: reward-to-go}
\hat{R}_t = r(t) + \gamma r(t + 1) + \ldots + \gamma^{T - t} r(T),
\end{equation}
which represents the cumulative discount sum of future rewards starting from time slot $t$ until the end of the trajectory at time slot $T$.
Finally, a batch of transitions are collected to train the networks. 
The actor networks are trained with the policy objective defined in Eq. \eqref{eq:actor}.  
The critic networks are trained with the loss objective defined in Eq. \eqref{eq:critic}. 
We optimize the policy objective and loss objective by Adam optimizer.

We enhance the policy using a PPO-clip \cite{schulman2017proximal} approach, ensuring that the parameter updates remain within conservative bounds.
Each agent's actor network optimizes the control policy to enhance the reward, updated by maximizing the following objective,
\begin{equation}\label{eq:actor}
\begin{aligned}
L(\theta) =\frac{1}{B}\sum_{i=1}^{B}[min(\eta_{i,\theta}\hat{A}_{i},clip(\eta_{i,\theta},1-\epsilon,1+\epsilon)\hat{A}_{i})\\ + \sigma S(\mu_{\theta}(o_{i}))],
\end{aligned}
\end{equation}
where $B$ is the batch size for buffer sampling, 
$\eta_{i,\theta} = \frac{\mu_{\theta}(a_{i}\mid o_{i})}{\mu_{\theta_{\text{old}}}(a_{i}\mid o_{i})} $ denotes the probability ratio with importance sampling, allowing the use of samples under parameters $\theta_{\text{old}}$ to update parameters $\theta$,
$\hat{A}_{i}$ is computed by GAE, evaluating the quality of the state-action pair ($\hat{A}_{i} > 0$ indicates relatively good; otherwise, relatively poor), 
$\epsilon$ is the hyperparameter controlling clipping strength to prevent significant changes in the new policy relative to the old policy, 
$S$ represents policy entropy to enhance exploration, %
and $\sigma$ is the coefficient of policy entropy.
We utilize gradient ascent with a learning rate of $\alpha$ and Adam optimization to update parameters $\theta'$ for maximizing the policy objective.

We represent the critic network of an agent as $V_{\phi}$ for GAE computation. The critic network is updated by minimizing the following loss,
\begin{equation}
\begin{aligned}
C(\phi) =  \frac{1}{B} \sum_{i=1}^{B}max[(V_{\phi}(s_{i})-\hat{R}_{i})^{2},( clip( V_{\phi}(s_{i}),\\ V_{\phi_{old}}(s_{i})-\varepsilon, V_{\phi_{old}}(s_{i})+\varepsilon )-\hat{R}_{i})^{2}],
\end{aligned}
\label{eq:critic}
\end{equation}
where $ \hat{R}_{i} $ represents the discounted reward-to-go, $s_i$ corresponds to the global state extracted from the buffer, $ V{\phi} $ denotes the value function with parameters $ \phi $, $ V_{\phi_{old}} $ refers to the value function with parameters $ \phi_{old} $, and $ \varepsilon $ serves as the hyperparameter regulating the clipping strength of the value loss. This is crucial for constraining the extent of change in the new value function relative to the old value function.
We utilize gradient descent to update the parameters $\phi'$ with a learning rate $\alpha$ to minimize the loss by Adam.

After training, the actor networks of the agents can be deployed on their respective edge nodes. The complexity of making a control decision is determined by the actor network's complexity, which is typically very small. The agents make control decisions based solely on their own states, eliminating the need for communication costs among the agents.

\begin{algorithm}[!h]
    \caption{Training Algorithm}
    \label{alg:algo}
    \begin{algorithmic}[1]
        \STATE Initialize the parameters $ \theta_i$ for the actor $ \mu$ and $ \phi_i$ for the critic $ V $ for each agent
        \FOR{$ episode = 1,...,M $}
            \STATE Reset the system and set the trajectory of each agent $\tau_i = [], i = 1,2,...,N$
            \FOR{$ t = 1,...,T $}
                \STATE Each agent $i \in N$ gets action $a_i(t)$ by Eq. \eqref{eq: sample action}
                \STATE Each agent executes its control action in the video analytics system, $a_1(t), ..., a_N(t)$
                \STATE Obtain the shared reward $ r(t) $ by Eq. \eqref{eq: share reward}
                \STATE Obtain next state $o_i(t+1), s(t+1)$ 
                \STATE Each agent $i=1,2,...,N$ updates the trajectory:
                    \STATE \hspace{\algorithmicindent}  $ \tau_i = \tau_i \cup [o_{i}(t), s(t), a_{i}(t), r(t), o_{i}(t+1), s(t+1)] $
            
            \ENDFOR

                \STATE Each agent $i \in N$ computes advantage estimation $ \hat{A} $ by GAE on $ \tau_i $ by Eq. \eqref{eq: GAE}
            \STATE Compute reward-to-go $ \hat{R} $ on $ \tau_i $ by Eq. \eqref{eq: reward-to-go}
                \STATE Store $ [o_{i}(t), s(t), a_{i}(t), r(t), o_{i}(t+1), s(t+1), \hat{A}, \hat{R}]$ into the replay buffer.
                
                \FOR{mini batch $j = 1,...,J $}
            \STATE $ batch = $ randomly select a mini batch $B$ from replay buffer $D$
                \STATE Edge agent updates $\theta_i$ and $\phi_i$:
                    
                \STATE \hspace{\algorithmicindent} Apply gradient ascent on $\theta_i$ by Adam 
                \STATE \hspace{\algorithmicindent} Apply gradient descent on $\phi_i$ by Adam
        \ENDFOR
        \STATE Empty replay buffer $D$
\ENDFOR
    \end{algorithmic}
\end{algorithm}

\section{Performance Evaluation} \label{sec:experiment}
In this section, we illustrate the experimental settings and evaluate the performance of our proposed method.




%
\subsection{Experimental Setting}
\textbf{Testbed:}
We evaluate a multi-edge video analytics system comprising four edge nodes. Each edge node is equipped with an Inter(R) Xeon(R) Gold 6242R CPU and a GeForce RTX 2080Ti GPU. The operating system across all edge nodes is Ubuntu 18.04, and the video analytics system is implemented using Pytorch and Python.
We use the road traffic videos \cite{website:video} as test video contents.
To emulate bandwidth connections between the edge nodes, publicly available bandwidth traces are adopted \cite{akhtar2018oboe}. 
In the absence of publicly available datasets for inference request rates, we simulate the arrival rates of inference requests on different edge nodes by scaling the request rates of the Wikipedia website \cite{urdaneta2009wikipedia}. 
The request rates on one edge node are relatively light compared to its processing capacity, while two edge nodes experience moderate request rates, and one edge node bears a heavy workload.

\textbf{System configuration:}
We perform object detection for our case study in video analytics. Four DNN-based object detection models, consisting of two small models (fasterrcnn mobilenet 320 and fasterrcnn mobilenet \cite{ren2015faster}) and two large models (retinanet resnet-50 \cite{lin2017focal} and maskrcnn resnet-50 \cite{he2017mask}), are deployed on each edge node. The default penalty weight for the overall delay is 5. The original video resolution is 1080P and can be downsized to 720P, 480P, 360P, and 240P through video preprocessing.
The recognition accuracy and average inference delay for a video frame under various models and resolutions are presented in Tables \ref{tab:accuracy} and \ref{tab:delay}, respectively. 
To calculate the recognition accuracy under different video resolutions, we compare it with the ground truth obtained from the recognition results of the most accurate model under the original resolution, following the approach consistent with prior works \cite{xiao2021towards,jiang2018chameleon,kang2017noscope}.

\begin{table}[htbp]
\tiny
\centering
\caption{The accuracy under different configurations.}
\resizebox{\linewidth}{!}{
\begin{tabular}{|c|c|c|c|c|c|}
\hline
Model name&1080P&720P&480P&360P&240P  \\ 
\hline
fasterrcnn mobilenet 320
&0.4158&0.4056&0.3834&0.3795&0.3426\\
\hline
fasterrcnn mobilenet
&0.6503&0.6194&0.5987&0.5676&0.5055\\
\hline
retinanet resnet-50
&0.8202&0.7630&0.7341&0.6917&0.5858\\
\hline
maskrcnn resnet-50
&0.8614&0.8102&0.7807&0.7457&0.6191\\
\hline
\end{tabular}}
\label{tab:accuracy}
\end{table}

\begin{table}[htbp]
\tiny
\centering
\caption{The average delay under different configurations.}
\resizebox{\linewidth}{!}{
\begin{tabular}{|c|c|c|c|c|c|}
\hline
Model name&1080P&720P&480P&360P&240P  \\ 
\hline
fasterrcnn mobilenet 320
&0.087s&0.056s&0.037s&0.030s&0.026s\\
\hline
fasterrcnn mobilenet
&0.103s&0.065s&0.049s&0.045s&0.039s\\
\hline
retinanet resnet-50
&0.147s&0.113s&0.088s&0.074s&0.068s\\
\hline
maskrcnn resnet-50
&0.171s&0.138s&0.110s&0.090s&0.074s\\
\hline
\end{tabular}}
\label{tab:delay}
\end{table}

\textbf{Training setup:}
The MARL models are trained through 50,000 episodes, each comprising 100 time steps with a duration of 0.2 seconds per step. 
Both the actor and critic networks employ MLP architectures, consisting of two hidden layers of 128 neurons each. ReLU serves as the activation function, with LayerNorm applied to each hidden layer. The actor network outputs three categorical distributions representing available actions, while the critic network produces predicted values.
The embedding network consists of a single layer with 8 neurons, yielding 8 outputs. Additionally, the experiment utilizes a multi-head attention network with 8 heads. The learning rate is set to 0.0005, the entropy coefficient to $0.01$, and the clip hyperparameter to 0.2.

\textbf{Baseline methods:}
We compare our method's performance with the following baseline methods.

\emph{1) IPPO}: The edge nodes utilize Proximal Policy Optimization (PPO) \cite{schulman2017proximal} as a single-agent reinforcement learning method. Each edge node independently learns its policies, operating autonomously without collaboration among agents.

\emph{2) Local-PPO:} Each edge node processes inference requests locally and does not forward them to others. The selection of models and resolutions is performed using the PPO algorithm independently on each edge node.

\emph{3) Predictive:} The system controller makes decisions by minimizing the system cost in the next time slots with the predicted inference workloads in the next time slot.

\emph{4) Shortest-Queue:} The incoming inference requests during a time slot are directed to the edge node with the shortest waiting queue length. Concurrently, we explore two methods for model and resolution selection:
\begin{itemize}
    \item \emph{Min:} Opt for the smallest model and the lowest resolution.
    \item \emph{Max:} Opt for the largest model and highest resolution.
\end{itemize}

\emph{5) Random:} In a given time slot, incoming inference requests to an edge node are randomly distributed among available edge nodes. Model and resolution selection strategies include \emph{Min} and \emph{Max}.

\begin{figure}[h]
  \centering
  \includegraphics[width=0.9\linewidth]{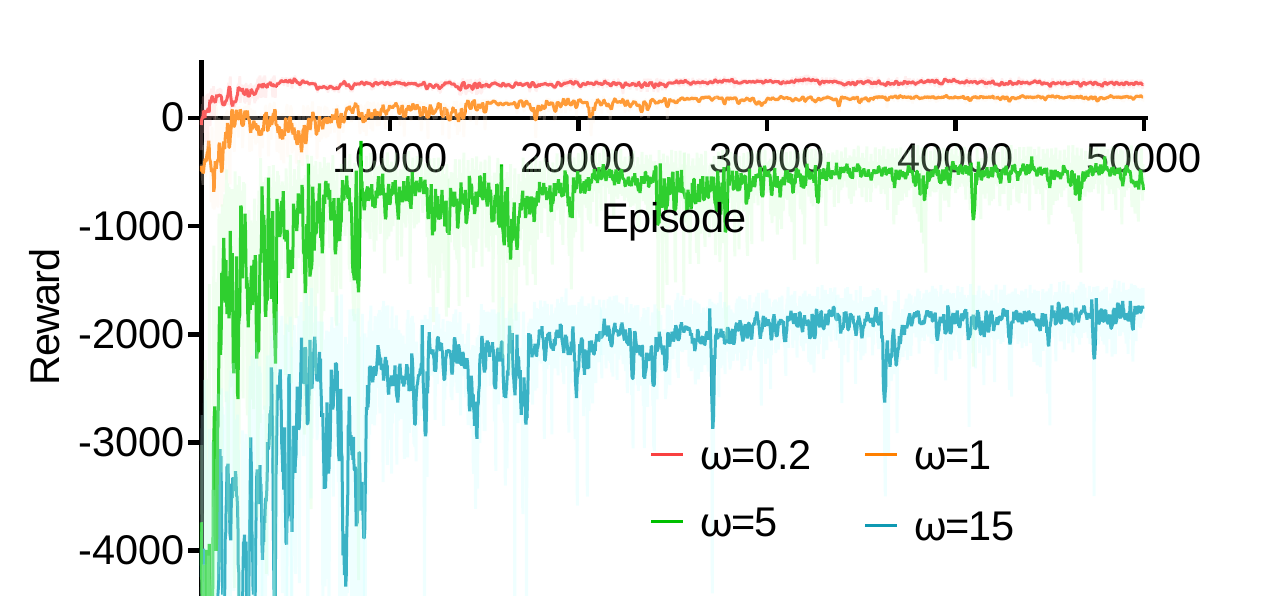}
  \caption{The convergence of our method under different penalty weights. Our algorithm can converge under different weights.}
  \label{fig: convergence}
\end{figure}

\subsection{Performance Analysis}
We first investigate the convergence of the training algorithm and then evaluate the performance characteristics of our method under varying penalty weights ($\omega$) for overall delay.

\textbf{Training convergence:}
In Fig. \ref{fig: convergence}, we evaluate the convergence of rewards using various penalty weights ($\omega=0.2,1,5,15$). Our method achieves convergence, enabling the learning of optimal policies across different weights. Notably, as the penalty weight for the overall delay increases, the converged rewards decrease, because larger weights impose greater penalties on the rewards for the same delay.

\begin{figure}[htbp]
\centering
\subfigure[Distributions of selected models.]{
\centering
\includegraphics[width=0.48\linewidth]{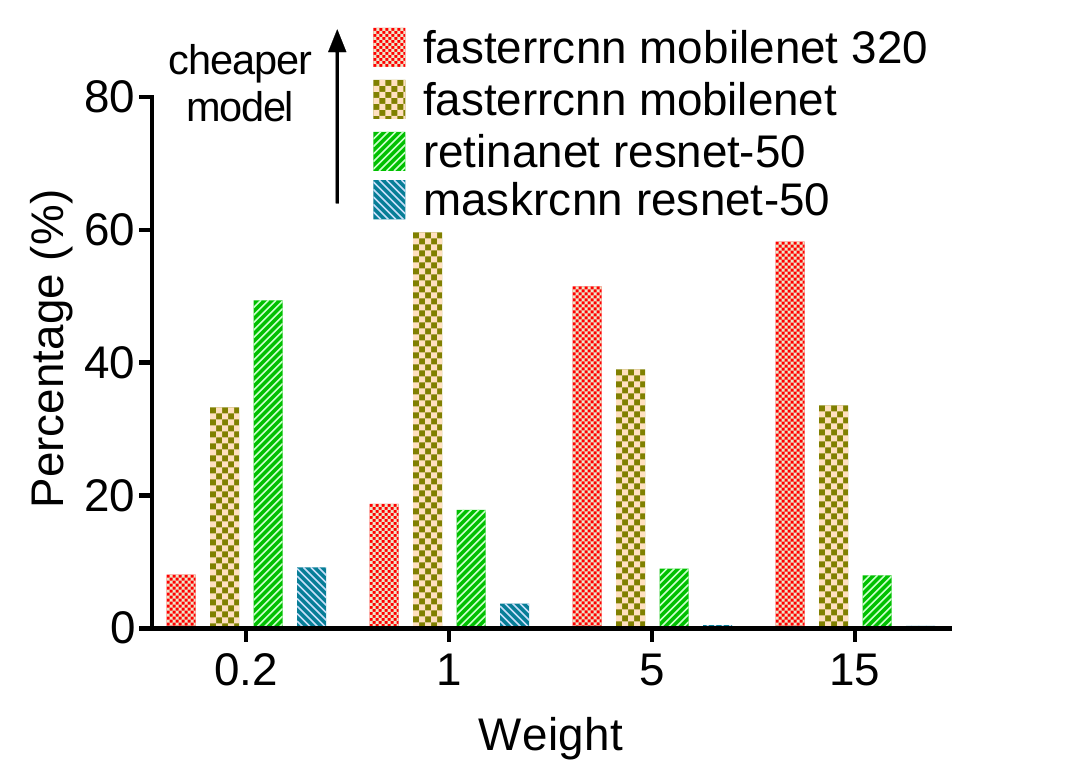} 
\label{fig: model}
}%
\subfigure[Distribution of selected resolutions.]{
\centering
\includegraphics[width=0.48\linewidth]{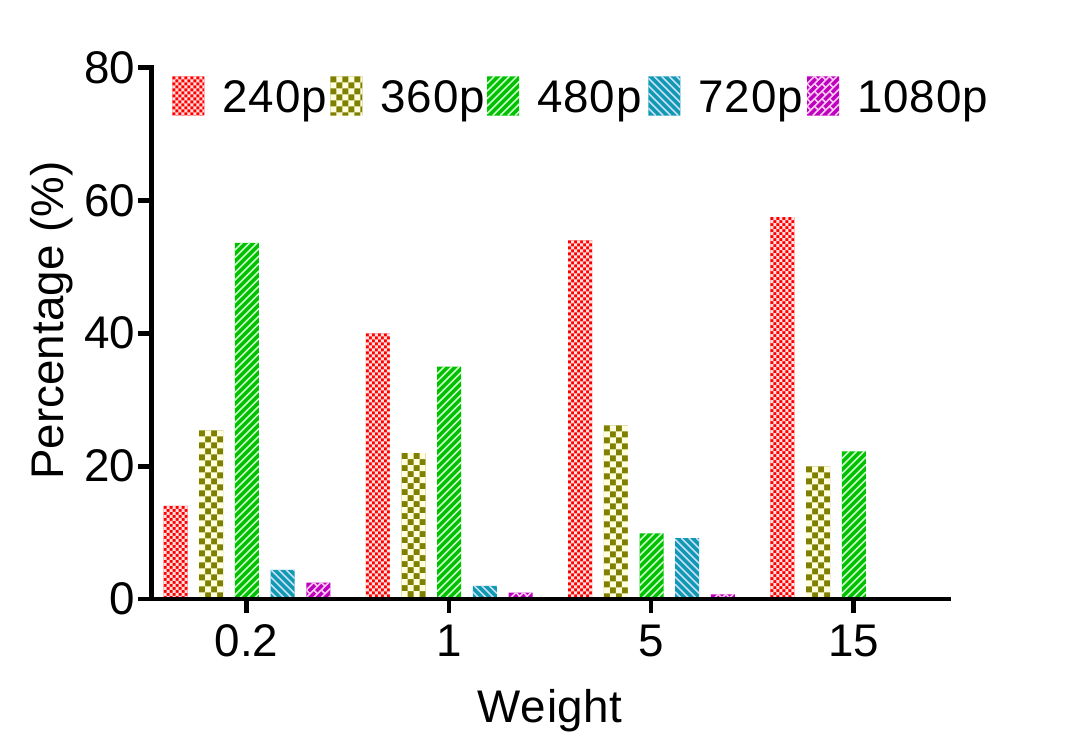}
\label{fig: resolution}
}%
\centering
\caption{The distributions of selected DNN modes for inference and selected resolutions for preprocessing under different weights.}
\label{fig: model and resolution}
\end{figure}

\begin{figure}[htbp]
\centering
\subfigure[Recognition accuracy.]{
\centering
\includegraphics[width=0.45\linewidth]{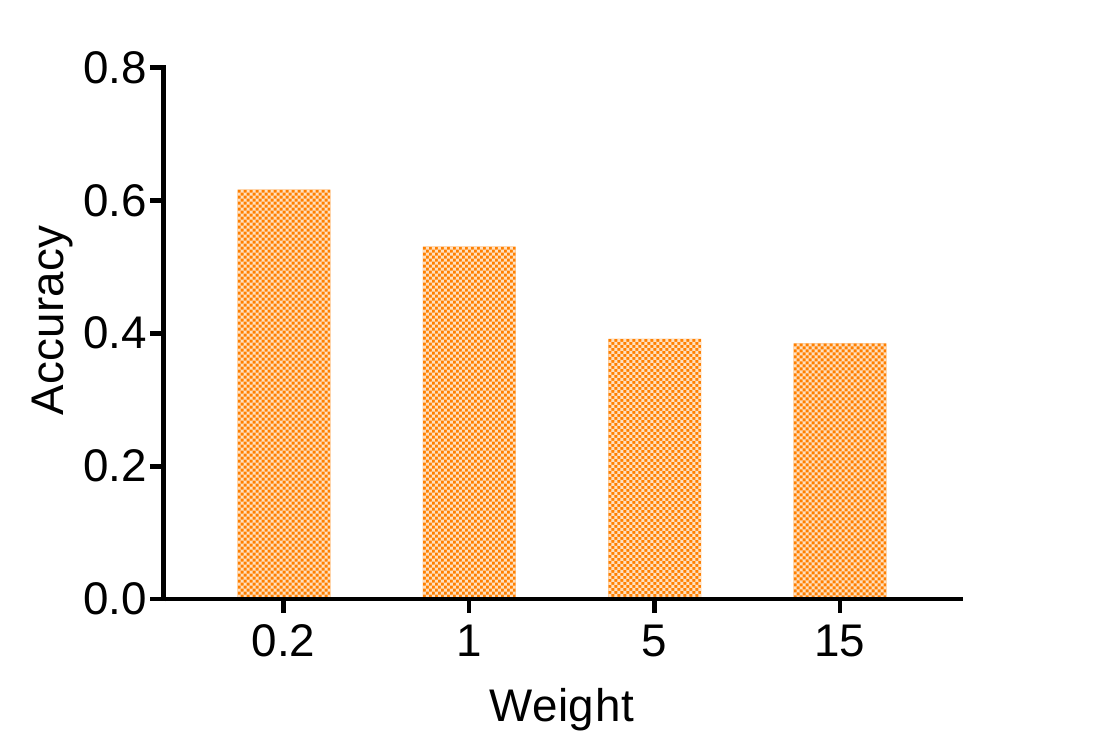} 
\label{fig: accuracy}
}%
\subfigure[Overall delay.]{
\centering
\includegraphics[width=0.45\linewidth]{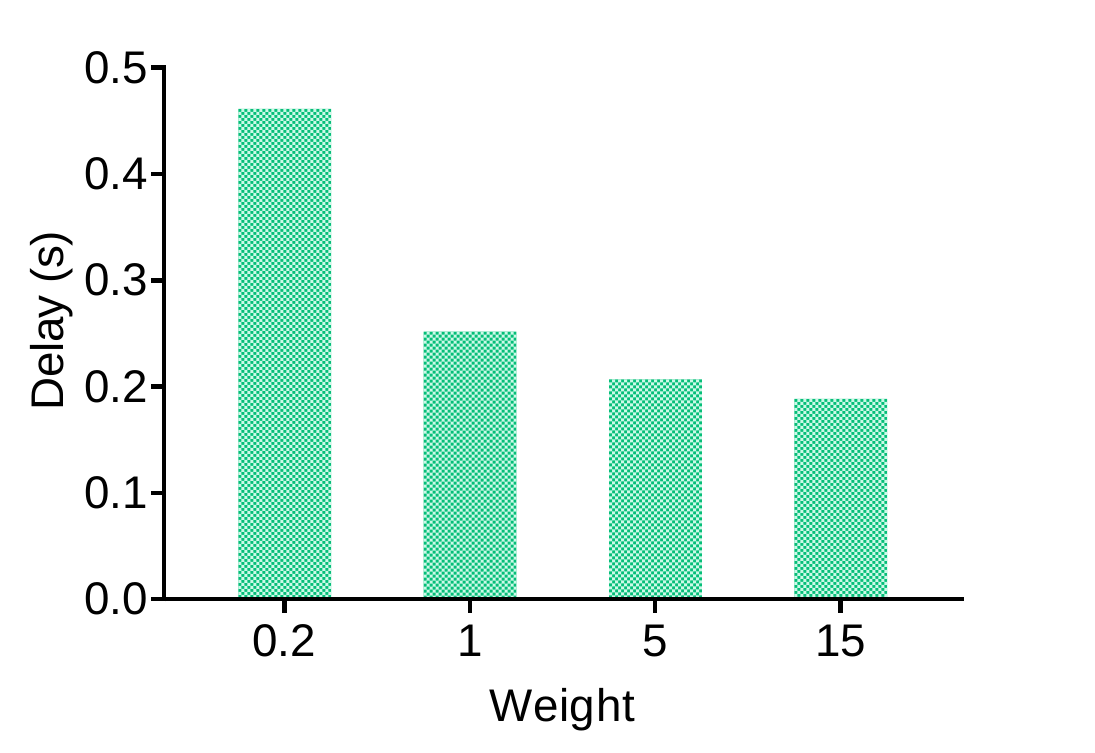}
\label{fig: delay}
}%

\subfigure[Dispatch percentage.]{
\centering
\includegraphics[width=0.45\linewidth]{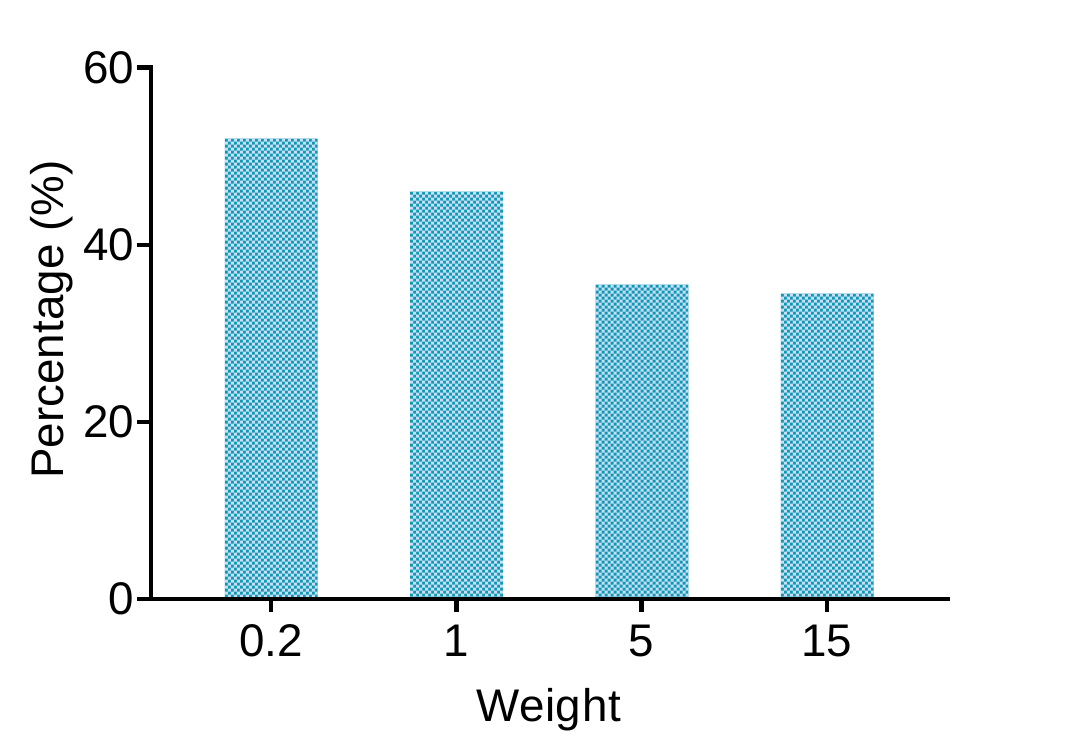}
\label{fig: dispatch}
}%
\subfigure[Frame drop percentage.]{
\centering
\includegraphics[width=0.45\linewidth]{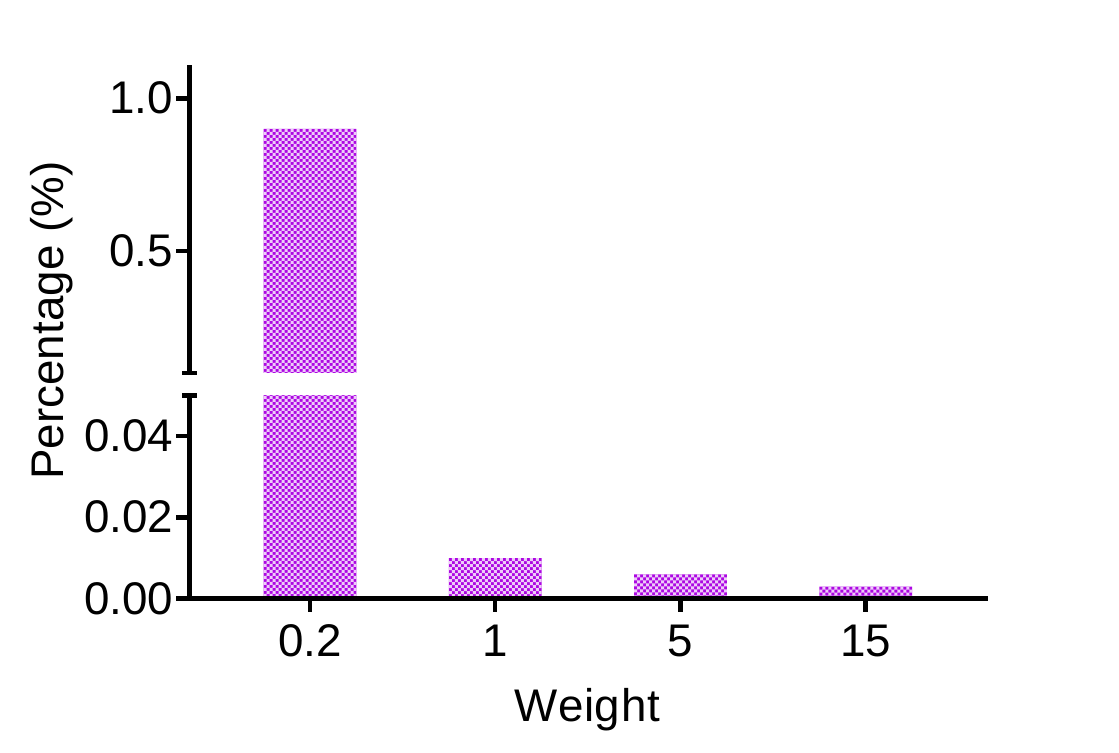}
\label{fig: drop}
}%

\centering
\caption{The average accuracy, delay, request dispatching percentage, and video frame drop percentage under different weights.}
\label{fig: accuracy delay dispatch and drop}
\end{figure}

\textbf{Performance characteristic:} 
In Figs. \ref{fig: model and resolution} and \ref{fig: accuracy delay dispatch and drop}, we examine the performance characteristics of our method across varying weights. These figures depict distributions of selected models and resolutions, average accuracy and overall delay per video frame, average request dispatching percentage, and average video frame drop percentage.

\emph{Model distribution:}
The distribution of chosen DNN models for edge node inference, illustrated in Fig. \ref{fig: model}, reveals that the percentage of large models (e.g., Mask R-CNN ResNet-50 and RetinaNet ResNet-50) decreases as weight increases. Conversely, the percentage of small models (e.g., Faster R-CNN MobileNet 320) increases. This shift is driven by the desire to minimize inference delay with cheaper models, resulting in higher rewards under larger penalty weights.

\emph{Resolution distribution:}
Fig. \ref{fig: resolution} illustrates the distribution of selected video resolutions for preprocessing on edge nodes under varying weights. The percentage of high-resolution choices (e.g., 1080P) decreases with increasing weight, while the percentage of downsized video frames (e.g., 240P) rises. Downsizing frames to lower resolutions reduces both inference and transmission delay, optimizing rewards when delay is a critical factor.

\emph{Inference accuracy:}
Fig. \ref{fig: accuracy} depicts the average inference accuracy of edge nodes across different weights. As weight increases, the average accuracy decreases due to downsizing more video frames to lower resolutions and analyzing them with smaller models, leading to a trade-off between accuracy and delay under larger weights.

\emph{Overall delay:}
The average overall delay for a video frame, as shown in Fig. \ref{fig: delay}, decreases with higher weight. This is attributed to more video frames being downsized to lower resolutions and analyzed with smaller models, resulting in reduced transmission and inference delay with larger weights.

\emph{Dispatching percentage:}
Fig. \ref{fig: dispatch} illustrates the average request dispatching percentage in the video analytics system for different weights. As the weight increases, the average dispatch percentage decreases, reflecting a preference for local analytics to reduce transmission delay.

\emph{Drop percentage:}
Fig. \ref{fig: drop} demonstrates the average video frame drop percentage across different weights. With increasing weight, the drop percentage decreases, driven by the adoption of smaller models and lower resolutions, effectively reducing inference and transmission delay to enhance system processing capacity.

\subsection{Performance Comparison}

\begin{figure*}[htbp]
\centering
\subfigure[$\omega=0.2$.]{
\centering
\includegraphics[width=0.35\linewidth]{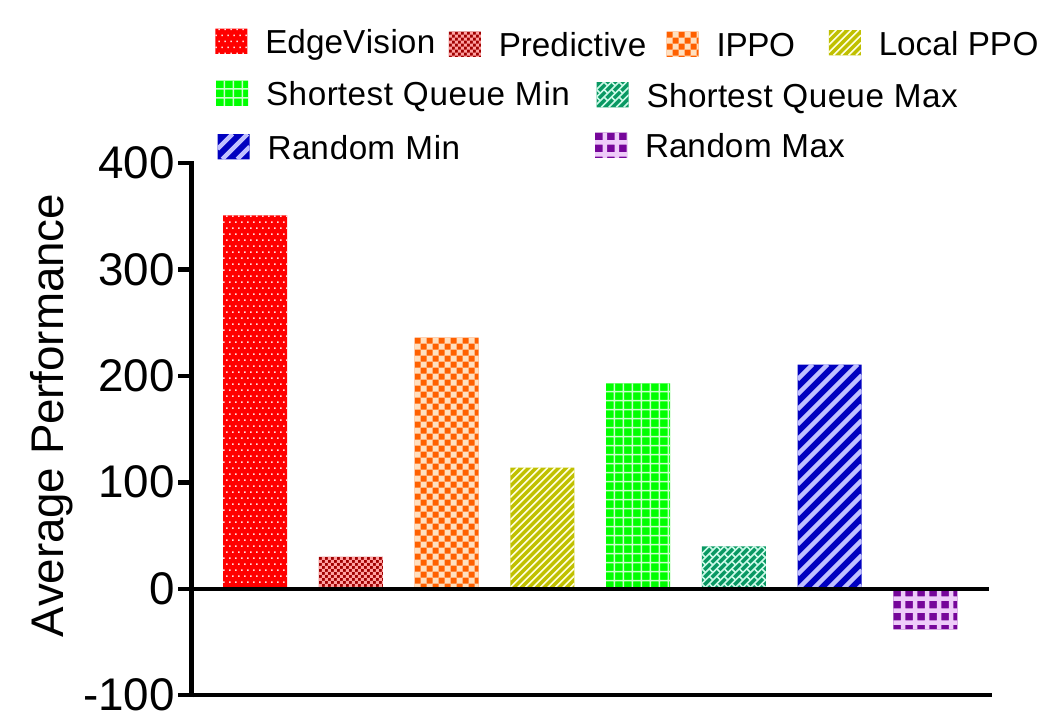} 
\label{fig:0.2}
}%
\subfigure[$\omega=1$.]{
\centering
\includegraphics[width=0.35\linewidth]{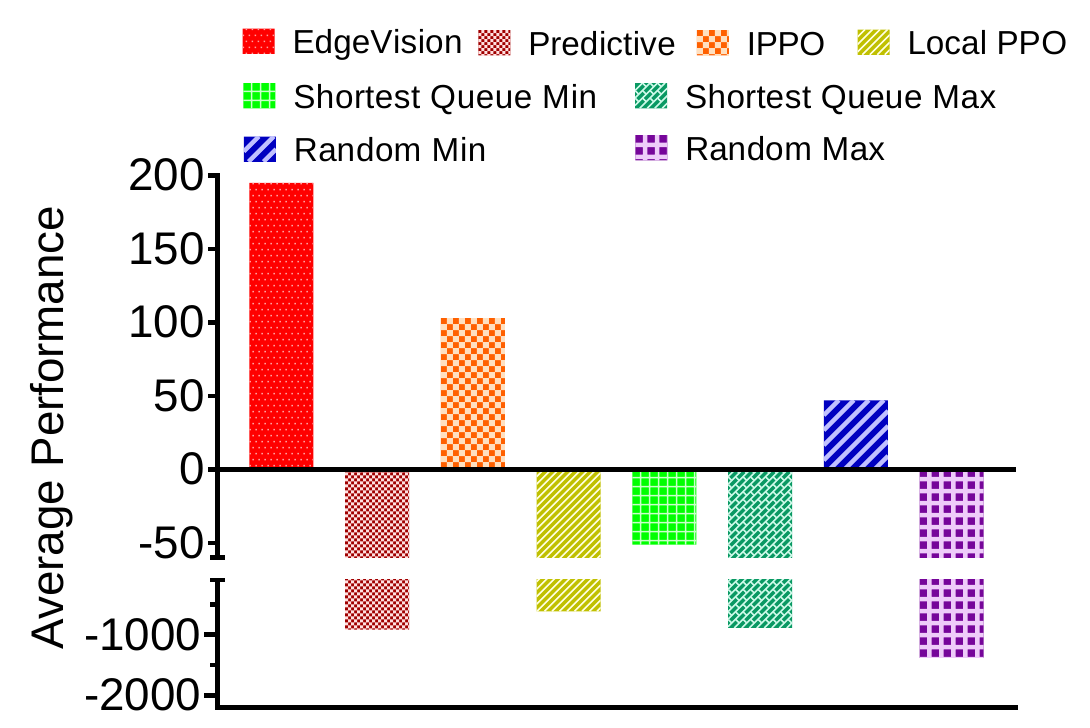}
\label{fig:1}
}
\subfigure[$\omega=5$.]{
\centering
\includegraphics[width=0.35\linewidth]{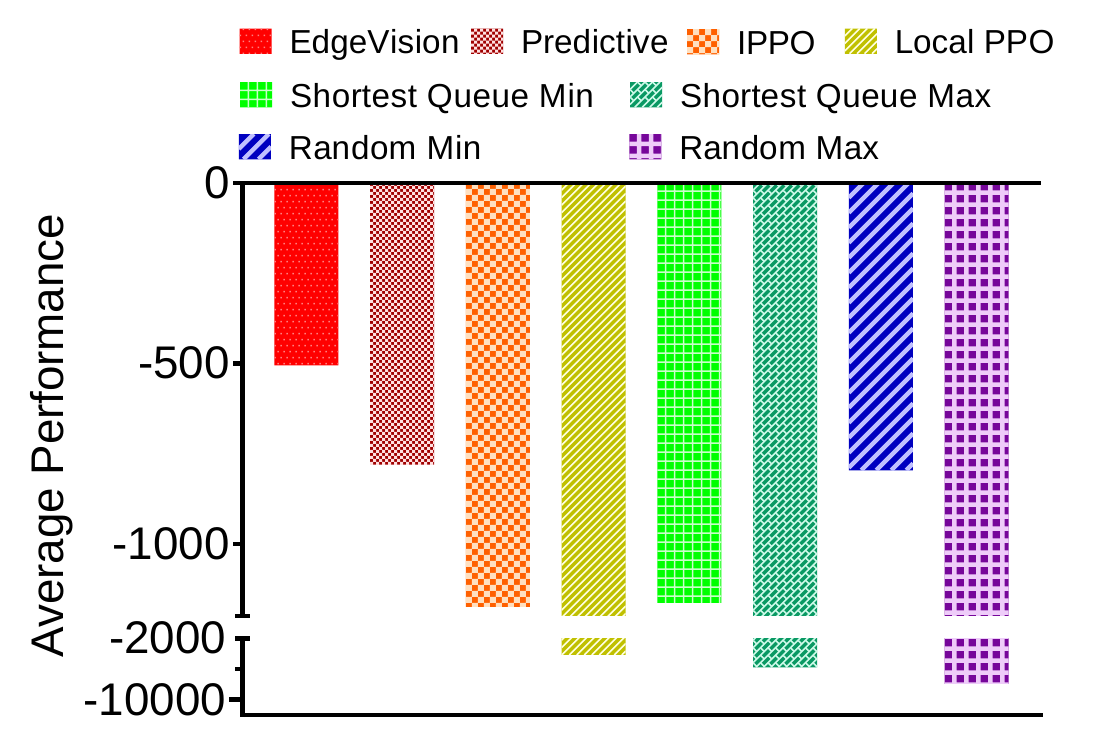} 
\label{fig:5}
}%
\subfigure[$\omega=15$.]{
\centering
\includegraphics[width=0.35\linewidth]{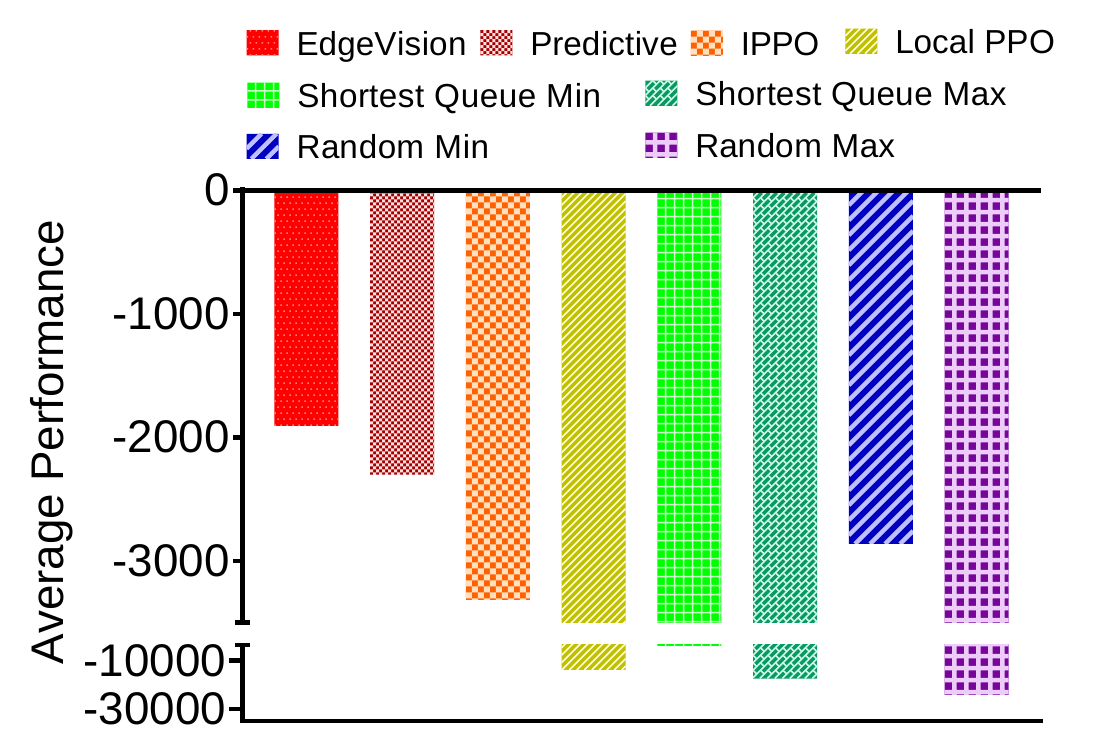}
\label{fig:15}
}%
\centering
\caption{The comparison of the average performance per episode of different methods under different penalty weights.}
\label{fig:cmp}
\end{figure*}

We present a comparative analysis of our method's average episode performance against baseline methods across various weights, as depicted in Fig. \ref{fig:cmp}. The results, illustrated in Fig. \ref{fig:cmp}, reveal that our method consistently outperforms baseline approaches across different weight configurations. Specifically, our method surpasses \emph{IPPO}, which exhibits lower rewards and increased instability during training due to independent policy learning for each agent. Meanwhile, \emph{Local-PPO} underperforms as edge nodes lack collaborative capabilities, limiting their ability to efficiently process video frames and handle inference requests collectively. \emph{Predictive} experiences diminished performance, as it may inaccurately anticipate future workloads and inadequately capture system dynamics.

In instances of relatively high penalty weights (i.e., $\omega = 5, 15$), baseline methods employing the most complex model and highest resolution (\emph{Shortest Queue Max}, \emph{Random Max}) demonstrate poor performance due to significant inference and transmission delays. Conversely, opting for the cheapest model and lowest resolution (\emph{Shortest Queue Min}, \emph{Random Min}) reduces overall delay and increases rewards. Nonetheless, our method consistently outperforms these alternatives by dynamically selecting the most suitable resolution, model, and edge node for video frames.

\begin{figure*}[htbp]
\centering
\subfigure[Average overall delay.]{
\centering
\includegraphics[width=0.3\linewidth]{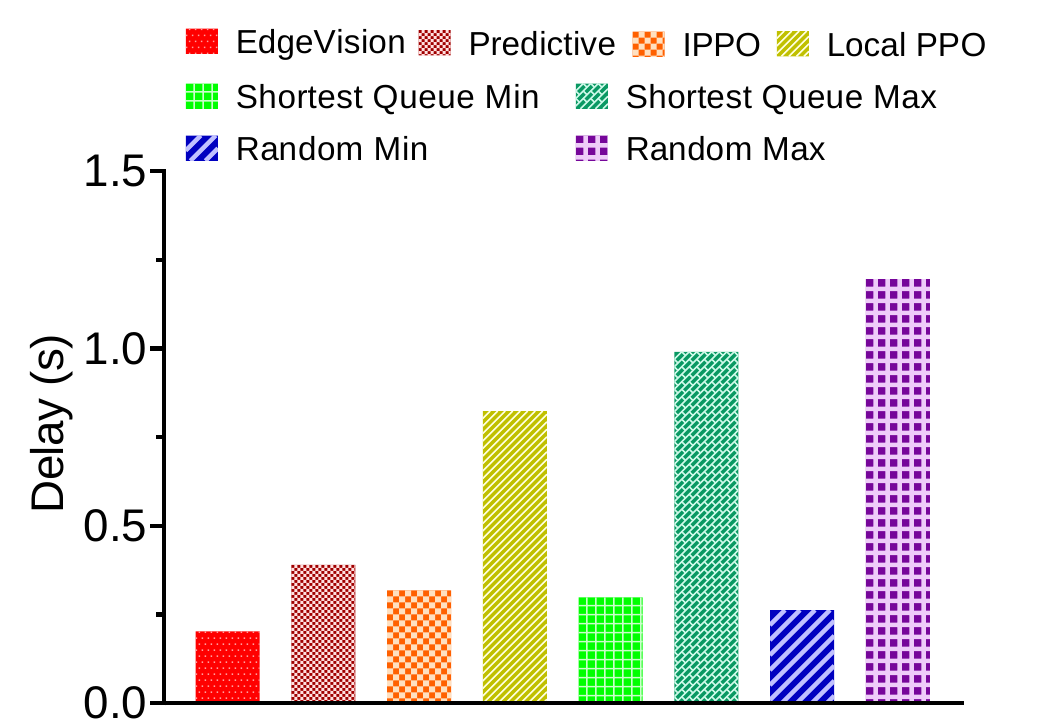}
\label{fig: w_5delay}
}
\subfigure[Average drop percentage.]{
\centering
\includegraphics[width=0.3\linewidth]{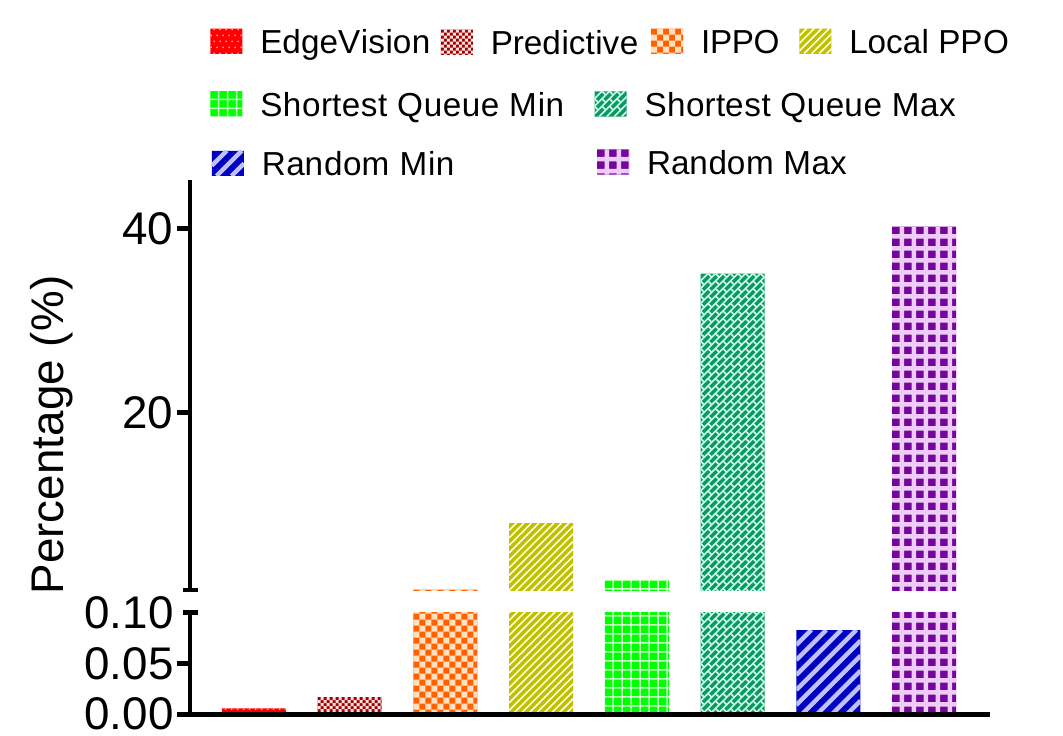}
\label{fig: w_5drop}
}%
\centering
\subfigure[Average inference accuracy.]{
\centering
\includegraphics[width=0.3\linewidth]{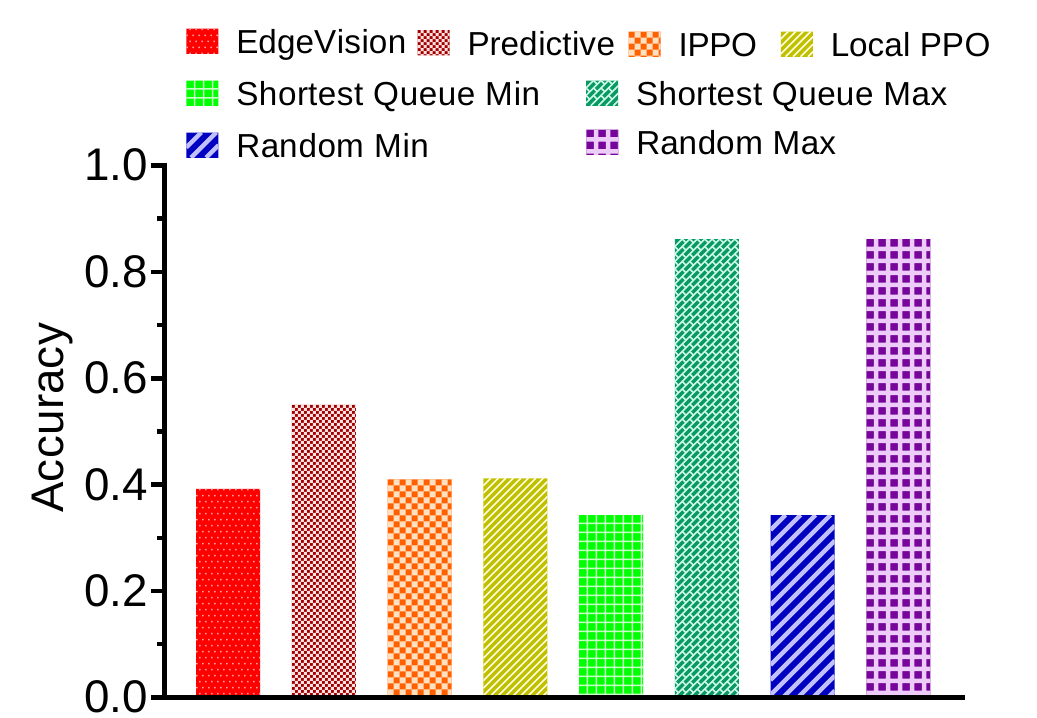} 
\label{fig: w_5accuracy}
}%
\caption{The comparison of the overall delay, video frame drop percentage, and average accuracy of different methods.}
\label{fig: cmp_w_5}
\end{figure*}

In Fig. \ref{fig: cmp_w_5}, we present a comprehensive analysis of various methods' average performance, focusing on metrics such as average accuracy, average overall delay, and average video frame drop percentage under the default penalty weight ($\omega = 5$). Similar observations hold for other penalty weights.
Fig. \ref{fig: w_5accuracy} shows that our method closely matches the accuracy of \emph{IPPO} and outperforms \emph{Predictive} in reducing overall delay and video frame drop percentage (Fig. \ref{fig: w_5delay} and \ref{fig: w_5drop}), highlighting the enhanced efficiency of our approach in scheduling inference requests across edge nodes.

In contrast, \emph{Local-PPO}, processing inference requests independently without dispatching, exhibits increased overall delay and video frame drop percentages in scenarios with high edge node workloads (Fig. \ref{fig: w_5delay} and \ref{fig: w_5drop}), indicating suboptimal performance. Baseline methods, \emph{Shortest Queue Min} and \emph{Random Min}, consistently select the cheapest model and the lowest resolution, resulting in lower accuracy and overall delay (Fig. \ref{fig: w_5accuracy} and \ref{fig: w_5delay}). However, their video frame drop percentages are higher compared to our method due to a lack of effective scheduling among edge nodes (Fig. \ref{fig: w_5drop}).

On the other hand, \emph{Shortest Queue Max} and \emph{Random Max}, consistently opting for the largest model and highest resolution, achieve the highest accuracy but suffer from increased overall delay and video frame drop percentages (Fig. \ref{fig: w_5accuracy}, \ref{fig: w_5delay}, and \ref{fig: w_5drop}). These methods exhibit poor performance, particularly under large penalty weights, due to significant overall delay despite the ability to dispatch inference requests among different edge nodes.

\begin{figure*}[htbp]
\centering
\subfigure[Average performance.]{
\centering
\includegraphics[width=0.35\linewidth]{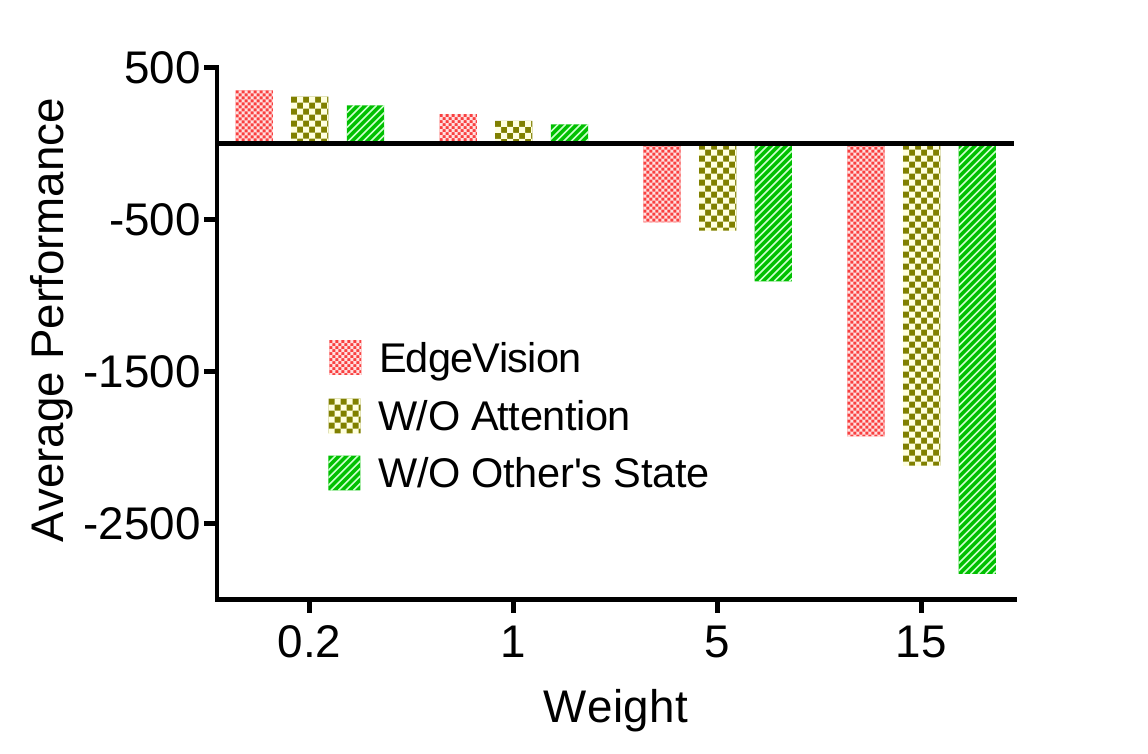} 
\label{fig: ablation reward}
}%
\subfigure[Average accuracy.]{
\centering
\includegraphics[width=0.35\linewidth]{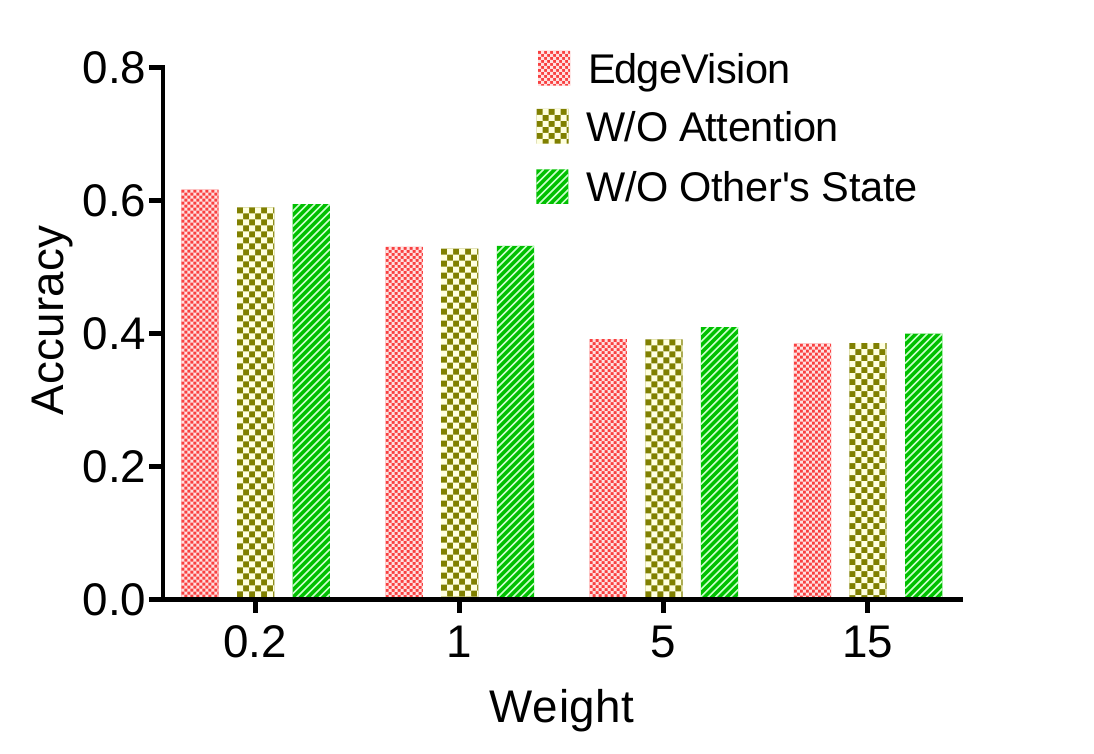}
\label{fig: ablation accuracy}
}
\subfigure[Average overall delay.]{
\centering
\includegraphics[width=0.35\linewidth]{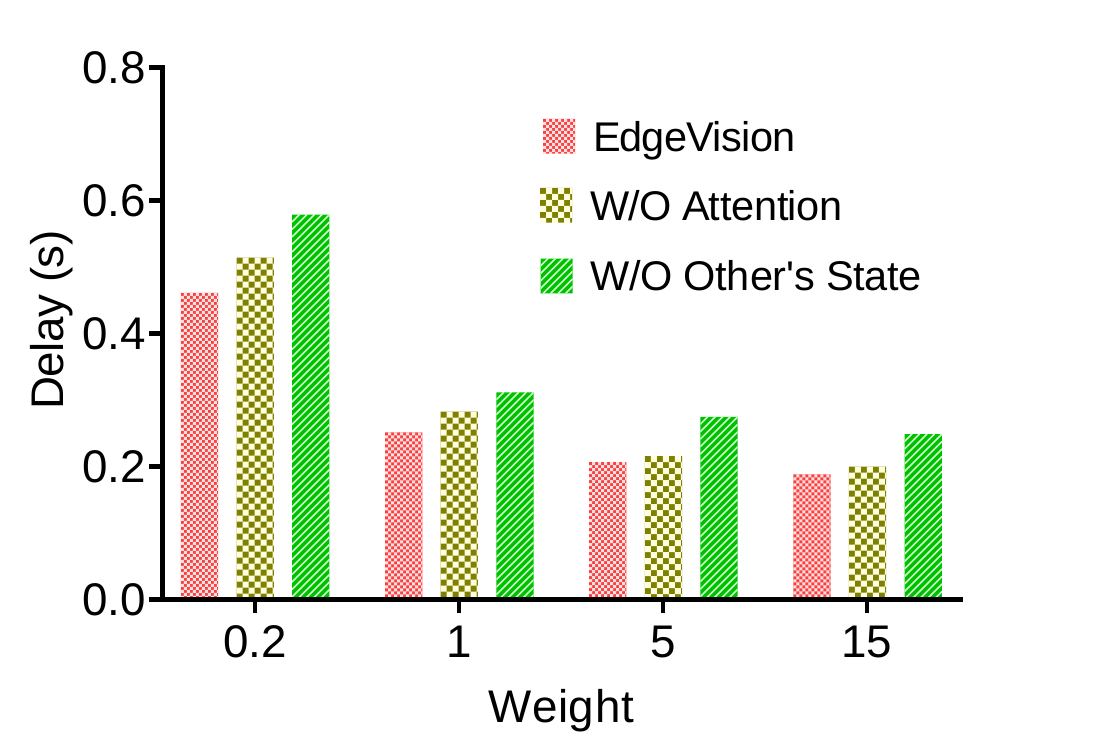} 
\label{fig: ablation delay}
}%
\subfigure[Average drop percentage.]{
\centering
\includegraphics[width=0.35\linewidth]{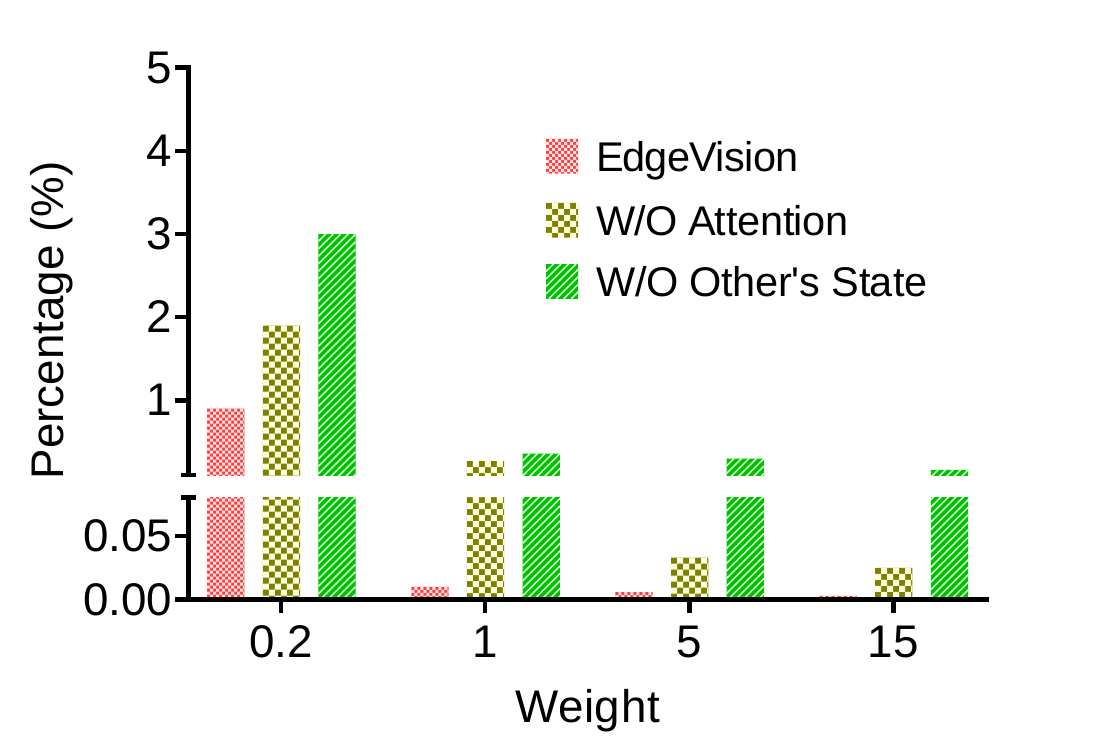}
\label{fig: ablation drop}
}%
\centering
\caption{The ablation study of our method under different penalty weights.}
\label{fig: ablation}
\end{figure*}

\subsection{Ablation Study}
In Fig. \ref{fig: ablation}, we perform an ablation study on our method, exploring the influence of varying penalty weights and dissecting the impact of its distinct components. \emph{W/O Attention} denotes our approach without the inclusion of the attention network, and \emph{W/O Other's State} characterizes our method with the exclusion of other agents' state information from the input of an agent's critic network.

We present the results in Fig. \ref{fig: ablation}, depicting average accuracy, overall delay, and video frame drop percentage across varying penalty weights per episode. Fig. \ref{fig: ablation reward} demonstrates superior performance of our approach compared to \emph{W/O Attention} and \emph{W/O Other's State} under different weights. \emph{W/O Other's State} exhibits the poorest performance, highlighting the challenge of agents learning optimal policies solely from local states. This underscores the necessity for the agent's critic network to incorporate information from other edge nodes during training for optimal policy learning. \emph{W/O Attention}, with undifferentiated attention to all agents' information, lags behind our method. The indiscriminate focus on some useless and uninformative states hampers training effectiveness, resulting in lower performance. Our approach outperforms \emph{W/O Attention} by 13.2\%, 30.0\%, 11.2\%, and 10.1\%, and surpasses \emph{W/O Other's State} by 38.7\%, 53.5\%, 42.9\%, and 32.8\% under weights of 0.2, 1, 5, and 15, respectively.

Fig. \ref{fig: ablation delay} and Fig. \ref{fig: ablation drop} demonstrate that our method achieves superior performance with the lowest overall delay and video frame drop percentage. This affirms its ability to efficiently schedule inference requests and make optimal selections of models and resolutions. For $\omega = 1$, Fig. \ref{fig: ablation accuracy} indicates comparable accuracy among the three methods, yet our approach outperforms others in terms of overall delay and video frame drop percentage, showcasing its effectiveness. At $\omega = 5$ and 15, Fig. \ref{fig: ablation accuracy} reveals that \emph{W/O Other's State} attains higher accuracy by opting for a more complex model and higher resolution. However, Fig. \ref{fig: ablation delay} and Fig. \ref{fig: ablation drop} unveil its drawbacks with higher overall delay and video frame drop percentage, leading to inferior overall performance.

\section{Conclusion} \label{sec:conclusion}
This paper introduced EdgeVision, a system for collaborative video analytics on distributed edges. Our approach is built on MARL, enabling edge nodes to collaborate during training and make independent distributed control decisions during execution. We leverage attention mechanisms to distill valuable information from large state spaces, enhancing learning capacity. In our system, edge nodes collaborate seamlessly for video frame preprocessing, model selection, and request dispatching to optimize overall performance. We implement a video analytics testbed with multiple edge nodes and conduct experiments using real-world datasets to evaluate our approach. Our experimental results demonstrate a significant improvement in overall system performance.

In this work, we assume that the computing capacities of the edge nodes are homogeneous, and each edge node is deployed with the same set of DNN models.
In our future work, we will address the challenge of heterogeneous computing capacities among edge nodes, allowing each node to dynamically choose the optimal set of DNN models for deployment under GPU memory constraints. 
Meanwhile, the DNN models deployed on each edge node are pre-trained. However, they may suffer from data drift issues, which will lead to the decrease of the recognition accuracy.
We will consider the continual learning scenarios for multi-edge collaborative video analytics.

\bibliographystyle{IEEEtran}
\bibliography{edge_system_ref}

\begin{thebibliography}{10}
\providecommand{\url}[1]{#1}
\csname url@samestyle\endcsname
\providecommand{\newblock}{\relax}
\providecommand{\bibinfo}[2]{#2}
\providecommand{\BIBentrySTDinterwordspacing}{\spaceskip=0pt\relax}
\providecommand{\BIBentryALTinterwordstretchfactor}{4}
\providecommand{\BIBentryALTinterwordspacing}{\spaceskip=\fontdimen2\font plus
\BIBentryALTinterwordstretchfactor\fontdimen3\font minus
  \fontdimen4\font\relax}
\providecommand{\BIBforeignlanguage}[2]{{%
\expandafter\ifx\csname l@#1\endcsname\relax
\typeout{** WARNING: IEEEtran.bst: No hyphenation pattern has been}%
\typeout{** loaded for the language `#1'. Using the pattern for}%
\typeout{** the default language instead.}%
\else
\language=\csname l@#1\endcsname
\fi
#2}}
\providecommand{\BIBdecl}{\relax}
\BIBdecl

\bibitem{hu2014picwords}
Z.~Hu, S.~Liu, J.~Jiang, R.~Hong, M.~Wang, and S.~Yan, ``Picwords: Render a
  picture by packing keywords,'' \emph{IEEE transactions on multimedia},
  vol.~16, no.~4, pp. 1156--1164, 2014.

\bibitem{xiao2021towards}
Z.~Xiao, Z.~Xia, H.~Zheng, B.~Y. Zhao, and J.~Jiang, ``Towards performance
  clarity of edge video analytics,'' in \emph{2021 IEEE/ACM Symposium on Edge
  Computing (SEC)}.\hskip 1em plus 0.5em minus 0.4em\relax IEEE, 2021, pp.
  148--164.

\bibitem{jiang2021joint}
J.~Jiang, Z.~Luo, C.~Hu, Z.~He, Z.~Wang, S.~Xia, and C.~Wu, ``Joint model and
  data adaptation for cloud inference serving,'' in \emph{2021 IEEE Real-Time
  Systems Symposium (RTSS)}.\hskip 1em plus 0.5em minus 0.4em\relax IEEE, 2021,
  pp. 279--289.

\bibitem{hao2022cdfkd}
Z.~Hao, Y.~Luo, Z.~Wang, H.~Hu, and J.~An, ``Cdfkd-mfs: Collaborative data-free
  knowledge distillation via multi-level feature sharing,'' \emph{IEEE
  Transactions on Multimedia}, vol.~24, pp. 4262--4274, 2022.

\bibitem{hu2018speeding}
Z.~Hu, P.~Sun, and Y.~Wen, ``Speeding-up age estimation in intelligent
  demographics system via network optimization,'' in \emph{2018 IEEE
  International Conference on Communications (ICC)}.\hskip 1em plus 0.5em minus
  0.4em\relax IEEE, 2018, pp. 1--7.

\bibitem{yi2017lavea}
S.~Yi, Z.~Hao, Q.~Zhang, Q.~Zhang, W.~Shi, and Q.~Li, ``Lavea: Latency-aware
  video analytics on edge computing platform,'' in \emph{Proceedings of the
  Second ACM/IEEE Symposium on Edge Computing}, 2017, pp. 1--13.

\bibitem{wu2023lean}
Q.~Wu, R.~Wang, X.~Duan, C.~Yi, and P.~Wang, ``A lean networking framework
  (leanet): Potential technical space and approaches for latency sensitive
  mobile services,'' \emph{IEEE Network}, 2023.

\bibitem{guan2020prefcache}
Y.~Guan, X.~Zhang, and Z.~Guo, ``Prefcache: Edge cache admission with user
  preference learning for video content distribution,'' \emph{IEEE Transactions
  on Circuits and Systems for Video Technology}, vol.~31, no.~4, pp.
  1618--1631, 2020.

\bibitem{cui2021towards}
L.~Cui, E.~Ni, Y.~Zhou, Z.~Wang, L.~Zhang, J.~Liu, and Y.~Xu, ``Towards
  real-time video caching at edge servers: A cost-aware deep {Q}-learning
  solution,'' \emph{IEEE Transactions on Multimedia}, 2021.

\bibitem{ma2017understanding}
G.~Ma, Z.~Wang, M.~Zhang, J.~Ye, M.~Chen, and W.~Zhu, ``Understanding
  performance of edge content caching for mobile video streaming,'' \emph{IEEE
  Journal on Selected Areas in Communications}, vol.~35, no.~5, pp. 1076--1089,
  2017.

\bibitem{qian2022osmoticgate}
B.~Qian, Z.~Wen, J.~Tang, Y.~Yuan, A.~Y. Zomaya, and R.~Ranjan, ``Osmoticgate:
  Adaptive edge-based real-time video analytics for the internet of things,''
  \emph{IEEE Transactions on Computers}, 2022.

\bibitem{ran2018deepdecision}
X.~Ran, H.~Chen, X.~Zhu, Z.~Liu, and J.~Chen, ``Deepdecision: A mobile deep
  learning framework for edge video analytics,'' in \emph{IEEE Conference on
  Computer Communications}.\hskip 1em plus 0.5em minus 0.4em\relax IEEE, 2018,
  pp. 1421--1429.

\bibitem{hung2018videoedge}
C.-C. Hung, G.~Ananthanarayanan, P.~Bodik, L.~Golubchik, M.~Yu, P.~Bahl, and
  M.~Philipose, ``Videoedge: Processing camera streams using hierarchical
  clusters,'' in \emph{2018 IEEE/ACM Symposium on Edge Computing (SEC)}.\hskip
  1em plus 0.5em minus 0.4em\relax IEEE, 2018, pp. 115--131.

\bibitem{zhang2023device}
P.~Zhang, F.~Huang, D.~Wu, B.~Yang, Z.~Yang, and L.~Tan, ``Device-edge-cloud
  collaborative acceleration method towards occluded face recognition in
  high-traffic areas,'' \emph{IEEE Transactions on Multimedia}, 2023.

\bibitem{wu2021edge}
D.~Wu, R.~Bao, Z.~Li, H.~Wang, H.~Zhang, and R.~Wang, ``Edge-cloud
  collaboration enabled video service enhancement: A hybrid human-artificial
  intelligence scheme,'' \emph{IEEE Transactions on Multimedia}, vol.~23, pp.
  2208--2221, 2021.

\bibitem{hao2022multi}
Z.~Hao, G.~Xu, Y.~Luo, H.~Hu, J.~An, and S.~Mao, ``Multi-agent collaborative
  inference via dnn decoupling: Intermediate feature compression and edge
  learning,'' \emph{IEEE Transactions on Mobile Computing}, 2022.

\bibitem{li2020reducto}
Y.~Li, A.~Padmanabhan, P.~Zhao, Y.~Wang, G.~H. Xu, and R.~Netravali, ``Reducto:
  On-camera filtering for resource-efficient real-time video analytics,'' in
  \emph{ACM Special Interest Group on Data Communication on the applications,
  technologies, architectures, and protocols for computer communication
  (SIGCOMM)}, 2020, pp. 359--376.

\bibitem{li2021appealnet}
M.~Li, Y.~Li, Y.~Tian, L.~Jiang, and Q.~Xu, ``Appealnet: An efficient and
  highly-accurate edge/cloud collaborative architecture for dnn inference,'' in
  \emph{2021 58th ACM/IEEE Design Automation Conference (DAC)}.\hskip 1em plus
  0.5em minus 0.4em\relax IEEE, 2021, pp. 409--414.

\bibitem{wang2019bridging}
Y.~Wang, W.~Wang, J.~Zhang, J.~Jiang, and K.~Chen, ``Bridging the {Edge-Cloud}
  barrier for real-time advanced vision analytics,'' in \emph{11th USENIX
  Workshop on Hot Topics in Cloud Computing (HotCloud 19)}, 2019.

\bibitem{jiang2018mainstream}
A.~H. Jiang, D.~L.-K. Wong, C.~Canel, L.~Tang, I.~Misra, M.~Kaminsky, M.~A.
  Kozuch, P.~Pillai, D.~G. Andersen, and G.~R. Ganger, ``Mainstream: Dynamic
  stem-sharing for multi-tenant video processing,'' in \emph{2018 USENIX Annual
  Technical Conference (USENIX ATC 18)}, 2018, pp. 29--42.

\bibitem{jiang2018chameleon}
J.~Jiang, G.~Ananthanarayanan, P.~Bodik, S.~Sen, and I.~Stoica, ``Chameleon:
  scalable adaptation of video analytics,'' in \emph{Proceedings of the 2018
  Conference of the ACM Special Interest Group on Data Communication}, 2018,
  pp. 253--266.

\bibitem{zeng2020distream}
X.~Zeng, B.~Fang, H.~Shen, and M.~Zhang, ``Distream: scaling live video
  analytics with workload-adaptive distributed edge intelligence,'' in
  \emph{Proceedings of the 18th Conference on Embedded Networked Sensor
  Systems}, 2020, pp. 409--421.

\bibitem{jingzong2023cross}
L.~Jingzong, L.~Liu, H.~Xu, S.~Wu, and C.~J. Xue, ``Cross-camera inference on
  the constrained edge,'' in \emph{42th IEEE International Conference on
  Computer Communications (IEEE INFOCOM 2023)}, 2023.

\bibitem{xu2023devit}
G.~Xu, Z.~Hao, Y.~Luo, H.~Hu, J.~An, and S.~Mao, ``Devit: Decomposing vision
  transformers for collaborative inference in edge devices,'' \emph{IEEE
  Transactions on Mobile Computing}, 2023.

\bibitem{zhao2022edgeadaptor}
K.~Zhao, Z.~Zhou, X.~Chen, R.~Zhou, X.~Zhang, S.~Yu, and D.~Wu,
  ``{EdgeAdaptor}: Online configuration adaption, model selection and resource
  provisioning for edge {DNN} inference serving at scale,'' \emph{IEEE
  Transactions on Mobile Computing}, 2022.

\bibitem{tan2021deep}
T.~Tan and G.~Cao, ``Deep learning video analytics through edge computing and
  neural processing units on mobile devices,'' \emph{IEEE Transactions on
  Mobile Computing}, 2021.

\bibitem{wang2022dynamic}
X.~Wang, G.~Gao, X.~Wu, Y.~Lyu, and W.~Wu, ``Dynamic dnn model selection and
  inference off loading for video analytics with edge-cloud collaboration,'' in
  \emph{Proceedings of the 32nd Workshop on Network and Operating Systems
  Support for Digital Audio and Video}, 2022.

\bibitem{zhang2020decomposable}
Y.~Zhang, J.-H. Liu, C.-Y. Wang, and H.-Y. Wei, ``Decomposable intelligence on
  cloud-edge {IoT} framework for live video analytics,'' \emph{IEEE Internet of
  Things Journal}, vol.~7, no.~9, pp. 8860--8873, 2020.

\bibitem{zhang2021towards}
M.~Zhang, F.~Wang, Y.~Zhu, J.~Liu, and Z.~Wang, ``Towards cloud-edge
  collaborative online video analytics with fine-grained serverless
  pipelines,'' in \emph{Proceedings of the 12th ACM Multimedia Systems
  Conference}, 2021, pp. 80--93.

\bibitem{rong2021scheduling}
C.~Rong, J.~H. Wang, J.~Liu, J.~Wang, F.~Li, and X.~Huang, ``Scheduling massive
  camera streams to optimize large-scale live video analytics,'' \emph{IEEE/ACM
  Transactions on Networking}, vol.~30, no.~2, pp. 867--880, 2021.

\bibitem{long2017edge}
C.~Long, Y.~Cao, T.~Jiang, and Q.~Zhang, ``Edge computing framework for
  cooperative video processing in multimedia {IoT} systems,'' \emph{IEEE
  Transactions on Multimedia}, vol.~20, no.~5, pp. 1126--1139, 2017.

\bibitem{wang2020joint}
C.~Wang, S.~Zhang, Y.~Chen, Z.~Qian, J.~Wu, and M.~Xiao, ``Joint configuration
  adaptation and bandwidth allocation for edge-based real-time video
  analytics,'' in \emph{IEEE Conference on Computer Communications}.\hskip 1em
  plus 0.5em minus 0.4em\relax IEEE, 2020, pp. 257--266.

\bibitem{deng2020fedvision}
Y.~Deng, T.~Han, and N.~Ansari, ``{FedVision}: Federated video analytics with
  edge computing,'' \emph{IEEE Open Journal of the Computer Society}, vol.~1,
  pp. 62--72, 2020.

\bibitem{du2020server}
K.~Du, A.~Pervaiz, X.~Yuan, A.~Chowdhery, Q.~Zhang, H.~Hoffmann, and J.~Jiang,
  ``Server-driven video streaming for deep learning inference,'' in \emph{ACM
  Special Interest Group on Data Communication on the applications,
  technologies, architectures, and protocols for computer communication
  (SIGCOMM)}, 2020, pp. 557--570.

\bibitem{elgamal2020sieve}
T.~Elgamal, S.~Shi, V.~Gupta, R.~Jana, and K.~Nahrstedt, ``Sieve: Semantically
  encoded video analytics on edge and cloud,'' in \emph{2020 IEEE 40th
  International Conference on Distributed Computing Systems (ICDCS)}.\hskip 1em
  plus 0.5em minus 0.4em\relax IEEE, 2020, pp. 1383--1388.

\bibitem{canel2019scaling}
C.~Canel, T.~Kim, G.~Zhou, C.~Li, H.~Lim, D.~G. Andersen, M.~Kaminsky, and
  S.~Dulloor, ``Scaling video analytics on constrained edge nodes,''
  \emph{Proceedings of Machine Learning and Systems}, pp. 406--417, 2019.

\bibitem{zhang2017live}
H.~Zhang, G.~Ananthanarayanan, P.~Bodik, M.~Philipose, P.~Bahl, and M.~J.
  Freedman, ``Live video analytics at scale with approximation and
  {Delay-Tolerance},'' in \emph{14th USENIX Symposium on Networked Systems
  Design and Implementation (NSDI 17)}, 2017, pp. 377--392.

\bibitem{zhao2021reinforcement}
Y.~Zhao, M.~Dong, Y.~Wang, D.~Feng, Q.~Lv, R.~P. Dick, D.~Li, T.~Lu, N.~Gu, and
  L.~Shang, ``A reinforcement-learning-based energy-efficient framework for
  multi-task video analytics pipeline,'' \emph{IEEE Transactions on
  Multimedia}, vol.~24, pp. 2150--2163, 2021.

\bibitem{zhang2023crossvision}
L.~Zhang, J.~Xu, Z.~Lu, and L.~Song, ``Crossvision: Real-time on-camera video
  analysis via common roi load balancing,'' \emph{IEEE Transactions on Mobile
  Computing}, 2023.

\bibitem{zhang2020deepqoe}
H.~Zhang, L.~Dong, G.~Gao, H.~Hu, Y.~Wen, and K.~Guan, ``Deepqoe: A multimodal
  learning framework for video quality of experience (qoe) prediction,''
  \emph{IEEE Transactions on Multimedia}, vol.~22, no.~12, pp. 3210--3223,
  2020.

\bibitem{schulman2015high}
J.~Schulman, P.~Moritz, S.~Levine, M.~Jordan, and P.~Abbeel, ``High-dimensional
  continuous control using generalized advantage estimation,'' \emph{arXiv
  preprint arXiv:1506.02438}, 2015.

\bibitem{schulman2017proximal}
J.~Schulman, F.~Wolski, P.~Dhariwal, A.~Radford, and O.~Klimov, ``Proximal
  policy optimization algorithms,'' \emph{arXiv preprint arXiv:1707.06347},
  2017.

\bibitem{website:video}
J.~Utah, ``Traffic camera videos and dash camera videos,'' Mar 2018,
  \url{https://www.youtube.com/channel/UCBcVQr-07MH-p9e2kRTdB3A/videos}.

\bibitem{akhtar2018oboe}
Z.~Akhtar, Y.~S. Nam, R.~Govindan, S.~Rao, J.~Chen, E.~Katz-Bassett,
  B.~Ribeiro, J.~Zhan, and H.~Zhang, ``Oboe: Auto-tuning video abr algorithms
  to network conditions,'' in \emph{Proceedings of the 2018 Conference of the
  ACM Special Interest Group on Data Communication}, 2018, pp. 44--58.

\bibitem{urdaneta2009wikipedia}
G.~Urdaneta, G.~Pierre, and M.~Van~Steen, ``Wikipedia workload analysis for
  decentralized hosting,'' \emph{Computer Networks}, vol.~53, no.~11, pp.
  1830--1845, 2009.

\bibitem{ren2015faster}
S.~Ren, K.~He, R.~Girshick, and J.~Sun, ``{Faster R-CNN}: Towards real-time
  object detection with region proposal networks,'' \emph{Advances in neural
  information processing systems}, vol.~28, 2015.

\bibitem{lin2017focal}
T.-Y. Lin, P.~Goyal, R.~Girshick, K.~He, and P.~Doll{\'a}r, ``Focal loss for
  dense object detection,'' in \emph{Proceedings of the IEEE international
  conference on computer vision}, 2017, pp. 2980--2988.

\bibitem{he2017mask}
K.~He, G.~Gkioxari, P.~Doll{\'a}r, and R.~Girshick, ``{Mask R-CNN},'' in
  \emph{Proceedings of the IEEE international conference on computer vision},
  2017, pp. 2961--2969.

\bibitem{kang2017noscope}
D.~Kang, J.~Emmons, F.~Abuzaid, P.~Bailis, and M.~Zaharia, ``Noscope:
  optimizing neural network queries over video at scale,'' \emph{arXiv preprint
  arXiv:1703.02529}, 2017.

\end{thebibliography}

\begin{IEEEbiography}[{\includegraphics[width=1in,height=1.28in,clip,keepaspectratio]
{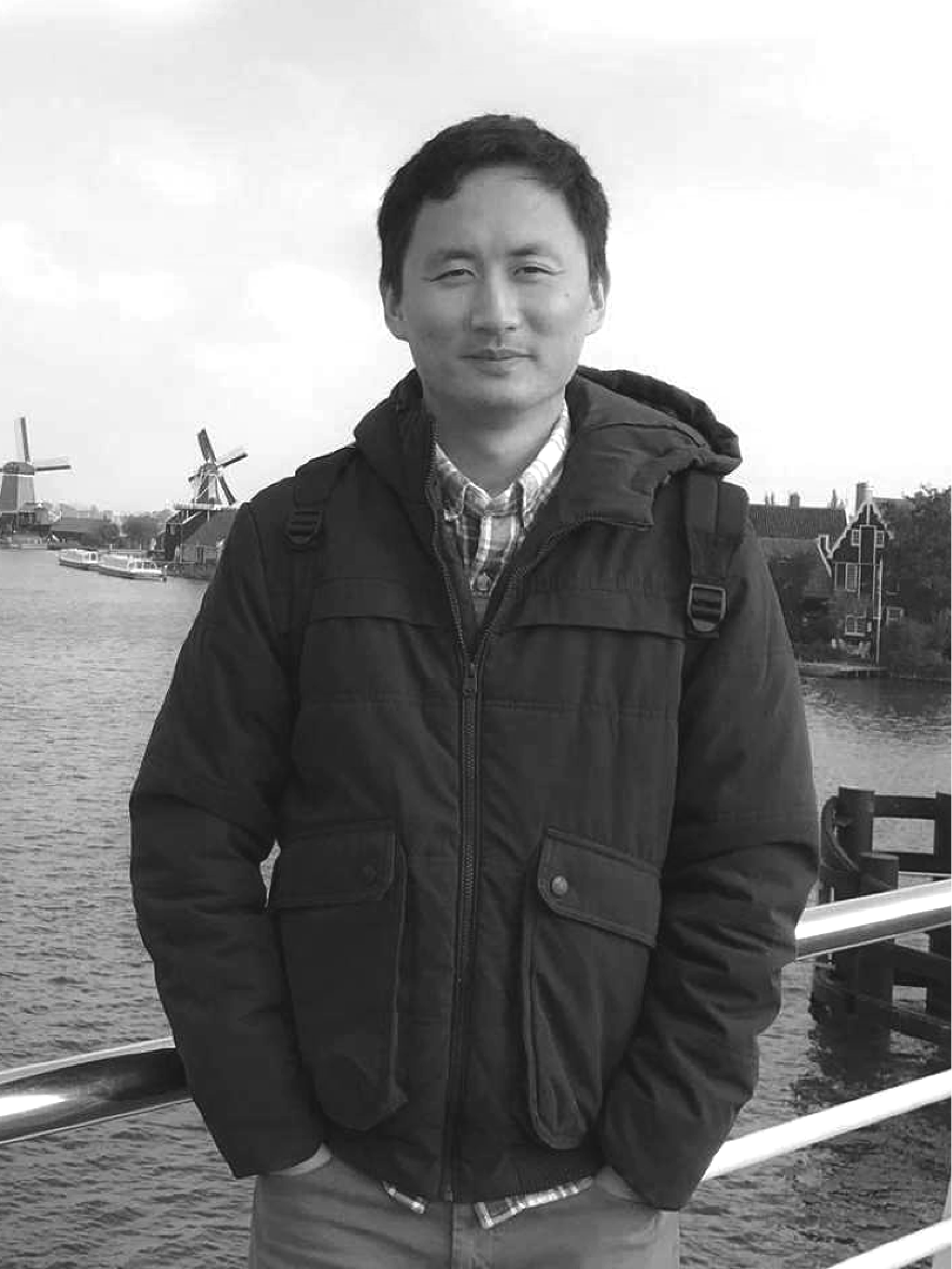}}]
{Guanyu Gao} is a Professor with the School of Computer Science and Engineering, Nanjing University of Science and Technology, Nanjing, China.
He received his Ph.D. degree from Nanyang Technological University, Singapore, in 2017, his M.S. degree from University of Science and Technology of China in 2012, and his B.S. degree from University of Electronic Science and Technology of China in 2009. 
He serves as the associate editor of IEEE Transactions on Network Science and Engineering. 
His research interests include multimedia networking, edge/cloud computing, edge intelligence, and internet of things.
\end{IEEEbiography}

\begin{IEEEbiography}[{\includegraphics[width=1in,height=1.28in,clip,keepaspectratio]
{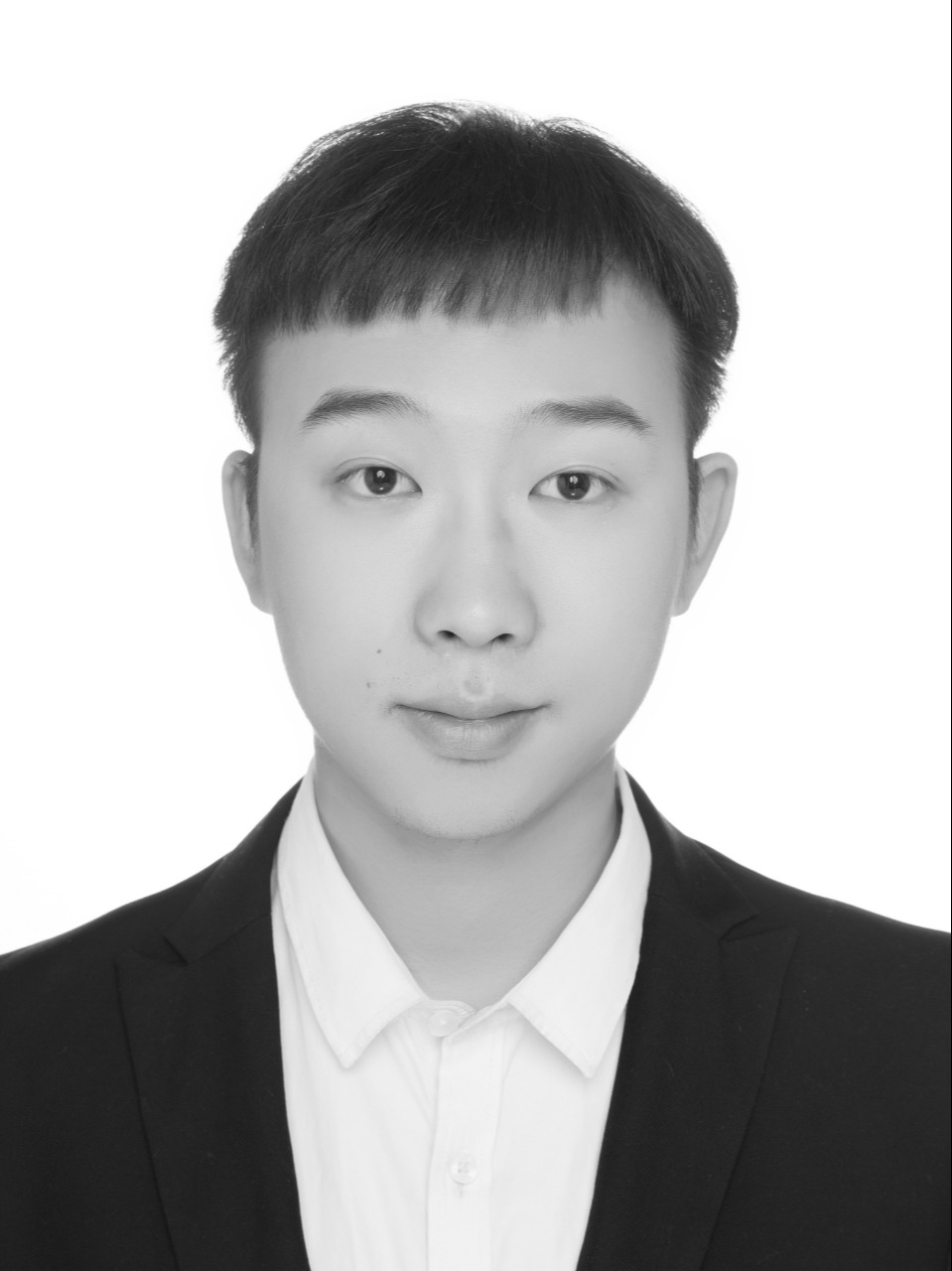}}]
{Yuqi Dong} received his M.S. degree from School of Computer Science and Engineering, Nanjing University of Science and Technology, Nanjing, China in 2024.
His research interests include edge/cloud computing, multi-agent reinforcement learning, machine learning inference, and video analytics.
\end{IEEEbiography}

\begin{IEEEbiography}[{\includegraphics[width=1in,height=1.28in,clip,keepaspectratio]
{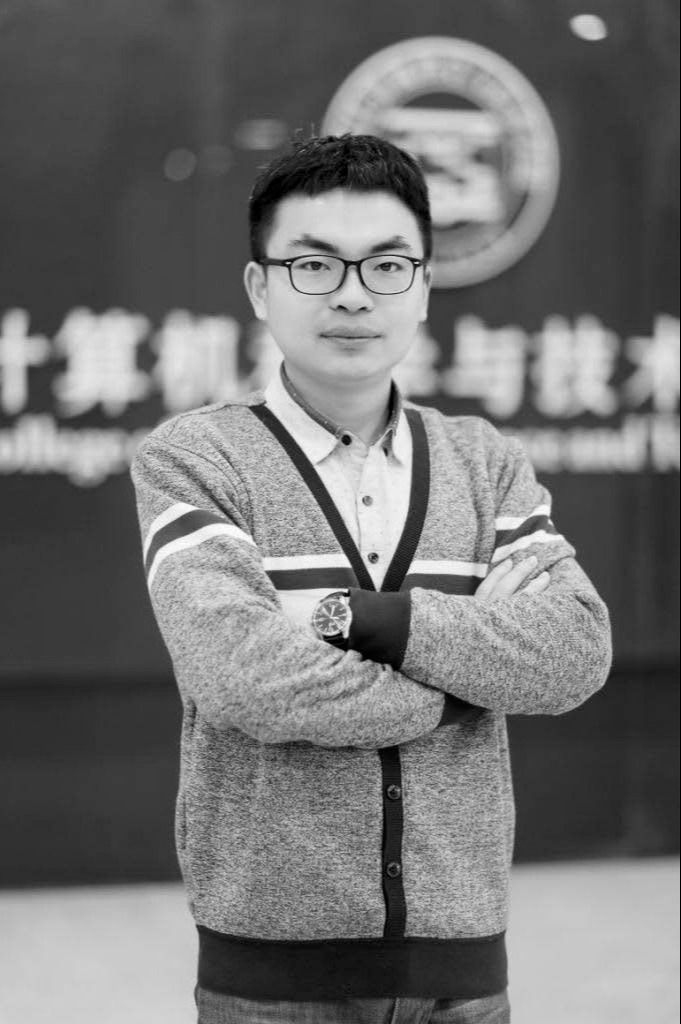}}]
{Ran Wang} (M'18) is an Associate professor and Doctoral Supervisor with the College of Computer Science and Technology, Nanjing University of Aeronautics and Astronautics, Nanjing, China. He received his B.E. in Electronic and Information Engineering from Honors School, Harbin Institute of Technology, China in July 2011 and Ph.D. in Computer Science and Engineering from Nanyang Technological University, Singapore in April 2016. He has authored or co-authored over 60 papers in top-tier journals and conferences. He received the Nanyang Engineering Doctoral Scholarship (NEDS) Award in Singapore and the innovative and entrepreneurial Ph.D. Award of Jiangsu Province, China in 2011 and 2017, respectively. He is the recipient of the Second Prize for Scientific and Technological Progress awarded by the China Institute of Communications. He is the Best Paper Award recipient of ACM MobiArch'23 and he is the ChangKong Scholar of NUAA. His current research interests include telecommunication networking and cloud computing.
\end{IEEEbiography}

\begin{IEEEbiography}[{\includegraphics[width=1in,height=1.28in,clip,keepaspectratio]
{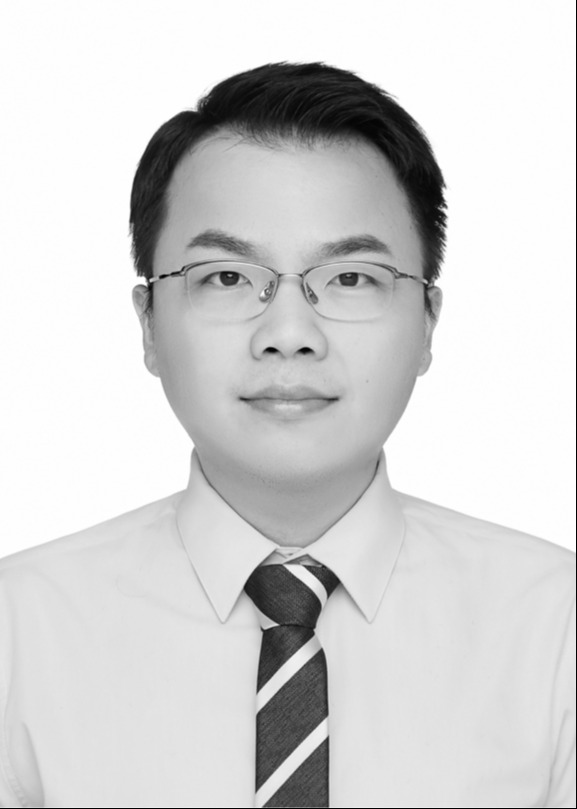}}]
{Xin Zhou} received M.E. and Ph.D. degrees from the Department of Information Engineering, Hiroshima University, Japan, in 2013 and 2016, respectively. He is now an Assistant Professor in the school of Information and Mechatronics Engineering, Jiangxi Science and Technology Normal University, China. His research interests include the learning-based optimization of ICT and cooling subsystems in the data center, reconfigurable architectures, parallel computing, parallel architecture, FPGA computing, deep reinforcement learning, and green data center. He received the Industrial Technical Excellence Award of the IEEE Technical Committee on Cyber–Physical Systems in 2020.
\end{IEEEbiography}

\end{document}